\documentclass[10pt,singlespace,doublecolumn]{IEEEtran}
\usepackage{cite}
\usepackage{amsmath}
\usepackage{amssymb}
\usepackage{graphicx}
\usepackage{epstopdf}
\usepackage{makecell}
\usepackage{color}
\usepackage{subfigure}
\usepackage{bm}
\usepackage{algorithmic,algorithm}
\usepackage{autobreak}
\usepackage{empheq} 
\usepackage[skins]{tcolorbox}
\newcounter{MYtempeqncnt}

\usepackage{orcidlink} 
\usepackage{hyperref}
\hypersetup{%
	colorlinks,
	citecolor=black,    
	linkcolor=black,   
	hypertexnames=false,
	hidelinks       
}

\let\oldcite=\cite
\renewcommand{\cite}[1]{\textcolor{black}{\oldcite{#1}}}

\begin{document}
\title{Fluid-Antenna-aided AAV Secure Communications in Eavesdropper Uncertain Location}
\author{Yingjie~Wu,
        Junshan~Luo,
        Weiyu~Chen,
        Shilian~Wang,\\
        Fanggang~Wang,~\IEEEmembership{Senior~Member,~IEEE},
        and Haiyang Ding,~\IEEEmembership{Member,~IEEE}
        \thanks{\scriptsize{Y. Wu, J. Luo, W. Chen, and S. Wang are with College of Electronic Science and Technology, National University of Defense Technology, Changsha 410073, China (email: wuyingjie@nudt.edu.cn, ljsnudt@foxmail.com, chenweiyu14@nudt.edu.cn, wangsl@nudt.edu.cn). 
        Corresponding author: Junshan~Luo and Shilian~Wang.} This work was supported by National Natural Science Foundation of China under Grant 62171445 and Grant 62201590.}
    	\thanks{\scriptsize{Fanggang Wang is with the School of Electronic and Information Engineering, Beijing Jiaotong University, Beijing 100044, China (e-mail:  wangfg@bjtu.edu.cn).}}
    	\thanks{\scriptsize{Haiyang Ding is with Engineering University of Information Support Force, Wuhan 430019, China (e-mail:dinghy2003@hotmail.com).}}
    	
}
\maketitle

\IEEEpeerreviewmaketitle

\begin{abstract}
For autonomous aerial vehicle (AAV) secure communications, traditional designs based on fixed position antenna (FPA) lack sufficient spatial degrees of freedom (DoF), which leaves the line-of-sight-dominated AAV links vulnerable to eavesdropping. To overcome this problem, this paper proposes a framework that effectively incorporates the fluid antenna (FA) and the artificial noise (AN) techniques. Specifically, the minimum secrecy rate (MSR) among multiple eavesdroppers is maximized by jointly optimizing AAV deployment, signal and AN precoders, and FA positions. In particular, the worst-case MSR is considered by taking the channel uncertainties due to the uncertainty about eavesdropping locations into account. To tackle the highly coupled optimization variables and the channel uncertainties in the formulated problem, an efficient and robust algorithm is proposed. Particularly, the uncertain regions of eavesdroppers, whose shapes can be arbitrary, are disposed by constructing convex hull.
In addition, two movement modes of FAs are considered, namely, free movement mode and zonal movement mode, for which different optimization techniques are applied, respectively. Numerical results show that, the proposed FA schemes boost security by exploiting additional spatial DoF rather than transmit power, while AN provides remarkable gains under high transmit power. Furthermore, the synergy between FA and AN results in a secure advantage that exceeds the sum of their individual contributions, achieving a balance between security and reliability under limited resources.


\end{abstract}
\begin{IEEEkeywords}
Fluid antenna, autonomous aerial vehicles communication, physical layer security, artificial noise, robust design.
\end{IEEEkeywords}

\section{Introduction}

\IEEEPARstart{A}{utonomous} aerial vehicles (AAVs), renowned for their high mobility, flexibility, and low cost, play a significant role in fulfilling the visions of sixth-generation (6G) wireless networks\cite{6GVTM,ASGMIN}. Their ability to dynamically establish line-of-sight (LoS) links makes them particularly valuable for expanding network coverage and enhancing service quality. For instance, AAVs can provide essential connectivity in the remote areas lacking terrestrial infrastructure, such as oceans, mountains, and deserts\cite{TWC22Fang}. In densely populated urban environments, they can serve as aerial base stations or mobile relays to boost capacity and enhance coverage\cite{TITS22Yu,JIOT20ji}.

Despite these advantages, AAV communications face critical security challenges\cite{TNSM24Khan}, which is one of the core requirements for 6G wireless communications. Compared to terrestrial systems, the LoS-dominated air-to-ground communications are more vulnerable to eavesdropping. Moreover, constraints on hardware resources, battery capacity, and computational capabilities often prevent AAVs from implementing complex cryptographic algorithms. Fortunately, physical-layer security, a complementary technology to traditional cryptography, can accomplish AAV secure communication based on the discrepancy of the wireless channels with no need of key distribution. The artificial noise (AN) and cooperative jammer are widely adopted physical-layer security technologies that can significantly enhance the achievable secrecy rate (SR)\cite{6JSYST22,8JIOT24,7TVT19,5TVT22,9TIFS24}. For instance, the authors in\cite{6JSYST22} investigated the trade-off between the security and transmission performance, and proposed an alternating optimization (AO) algorithm to design the AAV trajectory, power allocation, and user scheduling. Cao \textit{et al.}\cite{8JIOT24} proposed a novel AN scheme to improve security performance with minimal impact on reliability, where AN is used to encrypt the confidential signal as One-Time Pad. The work in\cite{7TVT19} considered a dual-AAV assisted secure communication system, where one AAV is for communication and the other AAV transmits the jamming signal to interfere with multiple eavesdroppers. The average of the minimum SR (MSR) over the flight period was maximized by jointly optimizing AAV trajectory and transmit power. Moreover, for a similar scenario, the work in\cite{5TVT22} considered the mobility of ground users as well as both the uplink and downlink transmission. A deep reinforcement learning algorithm was proposed to maximize the average MSR over the flight period. The authors in\cite{9TIFS24} studied the reconfigurable intelligent surface (RIS)-assisted AAV secure communication. For the scenarios where multiple eavesdroppers collude, the average SR was maximized by jointly optimizing the AAV trajectory, procoders of AN and signal, and RIS coefficients.

Since the channel state information (CSI) related to eavesdroppers is generally inaccurate or even unavailable, the robust secure AAV communications have been investigated\cite{1JSAC21,2TCOMM25,3TWC21,4TVT18,5TCOMM20}. In this field, Li \textit{et al.}\cite{1JSAC21} considered a statistical eavesdropper’s CSI (ECSI) error model, where the ECSI error is assumed to follow a complex Gaussian distribution. The secrecy energy efficiency of the considered dual-AAV-aided system was maximized. In contrast, the bounded CSI error model was considered in\cite{2TCOMM25} and\cite{3TWC21}. Specifically, the work in\cite{2TCOMM25} investigated the AAV secure communication of a cognitive network with the assistance of the interference from terrestrial BS. The minimum sum SR maximization problem was solved to achieve a win-win situation for the security of AAV user and the quality-of-service (QoS) of BS user. Different from above works, the authors in\cite{4TVT18} and\cite{5TCOMM20} considered the estimation errors about positions of eavesdroppers, where the worst-case SR was maximized by jointly optimizing the transmit power and the location of AAV. These studies underscore the importance of robust design in aerial secure communication systems.

Recently, the fluid antenna (FA), also known as movable antenna, has been recognized as a promising technology for enhancing the communication performance\cite{FA_Survey_COMST24,MA_Survey_MCOM23}. In contrast to the traditional fixed position antenna (FPA), FA can dynamically adjust the antenna positions and/or orientation within a confined space, thereby reshaping the wireless channel and exploiting additional spatial degrees of freedom (DoF). Notably, a limited number of FAs can achieve performance comparable to that of a larger FPA array\cite{BF_LCOMM24,4_Secrecy_TCOMM24}. Preliminary studies have demonstrated the potentials of FA in improving the channel capacity\cite{Capacity}, energy efficiency\cite{EE_TWC23}, and security performance\cite{1_Secrecy_GLOBE24,2_Secrecy_JIOT25,3_Secrecy_LSP24,4_Secrecy_TCOMM24,5_Secrecy_TMC24}. In the field of FA-assisted secure communications, the work in\cite{1_Secrecy_GLOBE24} addressed the SR maximization problem by discretizing the continuous antenna positions into multiple sampling positions and then by applying the graph theory. In\cite{2_Secrecy_JIOT25}, the author considered both the security and the covertness of communication, showing that FA can overcome the trade-off between improving the achievable secrecy rate and reducing the detection performance of warden. The works in\cite{3_Secrecy_LSP24} and\cite{4_Secrecy_TCOMM24} maximized the SR by jointly optimizing the transmit beamforming and FA positions. The results showed that FA can enhance the security performance by changing the correlation between different channels. Unlike the aforementioned studies, which assume perfect ECSI, in\cite{5_Secrecy_TMC24}, the secrecy outage probability was minimized merely relying on statistical ECSI. 

Motivated by the benefits of AAV and FA, their integration presents a highly promising avenue for advanced wireless communication systems\cite{AAV-FA_MWC24}. Existing studies have shown the advantages of employing FA for AAV communication systems in improving the capacity\cite{6AAV-FA}, reliability\cite{7AAV-FA}, and interference mitigation\cite{8AAV-FA}. For instance, the work in\cite{6AAV-FA} maximized the sum rate by jointly optimizing the beamforming, AAV trajectory, and FA position. In\cite{7AAV-FA}, the authors derived the outage probability of AAV-aided FA system, where AAV acts as an aerial relay. Ren \textit{et al.}\cite{8AAV-FA} equipped the AAV with FA to suppress the interference from non-associated BS. The results of the aforementioned research showed that the performances of FA-AAV systems significantly outperform those of FPA-AAV systems.

Despite these pioneering contributions \cite{6AAV-FA,7AAV-FA,8AAV-FA}, applying FA in AAV secure communications 	remains unexplored. Note that this integration is not a simple technical combination but a necessary exploration to address practical security demands. Specifically, traditional secure schemes for AAV communications mainly rely on FPA, whose performance is limited by the fixed correlation between different channels, and thus struggle to balance the security and reliability. FA technology has the potential to offer an innovative solution to break through this bottleneck. In particular, AAV offers large-scale channel reconfiguration through mobility, while FA provides small-scale adaptability through flexible antenna adjusting. These two scales of DoF complement each other, which may create a security barrier that FPA-based AAVs cannot achieve. Nonetheless, applying FA in AAV secure communications introduces some fundamentally new and intertwined challenges. On one hand, the security performance is jointly determined by the relative positions of AAV to receivers/eavesdroppers and the positions of FAs. How to jointly and fully exploit the large-scale channel reconfiguration ability of AAV and the small-scale adaptability of FA is worth investigating. On the other hand, in practice, the transmitter may only know the suspicious regions where eavesdroppers may exist, whose shapes may be irregular. This makes the security of the considered system not robust and makes it intractable to maximize the  worst-case MSR.
Motivated by these research gaps, this paper investigates the robust secure AAV communication with the joint assistance of FA and AN, considering practical imperfect ECSI conditions. Main contributions are summarized as below.

\begin{itemize}
	\item To address the critical challenges of securing AAV communications against multiple eavesdroppers with imperfect ECSI, a novel framework that synergizes FA technology and AN is proposed to overcome the limitations of conventional FPA systems. Our framework establishes a foundational advancement in security for aerial networks, as it can effectively tackle eavesdropping threats under uncertainty about locations of eavesdroppers, where FPA methods fail to provide sufficient spatial DoF.
	\item To rigorously address the uncertainties about locations of eavesdroppers, we formulate a worst-case MSR maximization problem, in which AAV deployment, signal and AN precoders, and antenna positions are jointly optimized. A convex hull is constructed to transform eavesdropper position uncertainties, converting location errors into tractable convex approximations that capture worst-case eavesdropping scenarios. This approach transcends conventional statistical models by deterministically bounding adversarial advantages under imperfect CSI, thereby enabling robust resource allocation without prior knowledge of eavesdropper distribution.
	\item We consider two movement modes for the FA, i.e., the free movement mode (FMM) and the zonal movement mode (ZMM), and propose two efficient algorithms for both modes. Specifically, for the FMM, the successive convex approximation (SCA) algorithm is adopted, while the alternating direction method of multipliers (ADMM) is used to address the ZMM. The results demonstrate universal security enhancement. In particular, the FA boosts SR at all power levels via additional spatial DoF, while AN provides supplementary confidentiality gains under sufficient transmit power.
\end{itemize}

The rest of this paper is organized as follows. In Section \ref{Section2}, the FA-aided AAV system model is introduced and the corresponding constrained optimization problem is formulated. Section \ref{Section 3} presents a robust scheme for jointly optimizing the four key system parameters. Section \ref{Section4} and Section \ref{Section5} provide discussions of numerical results and conclusions, respectively.

\textit{Notations}: $(\cdot)^\mathrm{T}$, $(\cdot)^\mathrm{H}$, and $(\cdot)^*$ denote the transpose, Hermitian transpose, and conjugate, respectively. $\mathrm{tr}(\boldsymbol{X})$ and $\mathrm{rank}(\boldsymbol{X})$ stand for the trace and rank of matrix $\boldsymbol{X}$. $\mathrm{diag}(\boldsymbol{x})$ denotes a diagonal matrix with each diagonal element given by vector $\boldsymbol{x}$. $\boldsymbol{I}_N$ denotes the $N\times N$ identity matrix. $\boldsymbol{e}_i$ is the $ i $-th column of a identity matrix. $ \otimes $ indicates the Kronecker product. $\left|\cdot\right|$, $\left\|\cdot\right\|_2$, and $\left\|\cdot\right\|_\mathrm{F}$ denote determinant, Euclidean norm, and Frobenius norm operations, respectively. $\mathbb{R}^{m \times n}$ and $\mathbb{C}^{m \times n}$ represent the space of $m \times n$ real and complex matrices, respectively. $\mathcal{I}_N\triangleq\{1,2,...,N\}$ denotes the set of integers from 1 to $ N $. $\nabla_{{{\boldsymbol{x}}}}f$ and $\nabla_{{{\boldsymbol{x}}}}^2f$
denote the gradient vector and Hessian matrix of function $f(\boldsymbol{x})$ with respect to (w.r.t.) variable $\boldsymbol{x}$. The distribution of a circularly symmetric complex Gaussian random vector with mean vector $\boldsymbol{\mu}$ and covariance matrix $\boldsymbol{\Sigma}$ is denoted by $\mathcal{CN}(\boldsymbol{\mu},\boldsymbol{\Sigma})$. $[\cdot]^+$ is the projection onto the non-negative value, i.e., $[x]^+\triangleq\max[x,0]$. $\mathcal{O}(\cdot)$ is for the standard big-O notation.

\section{System Model and Problem Formulation}
\label{Section2}
\begin{figure}[t!]
	\centering
	\includegraphics[scale=0.4]{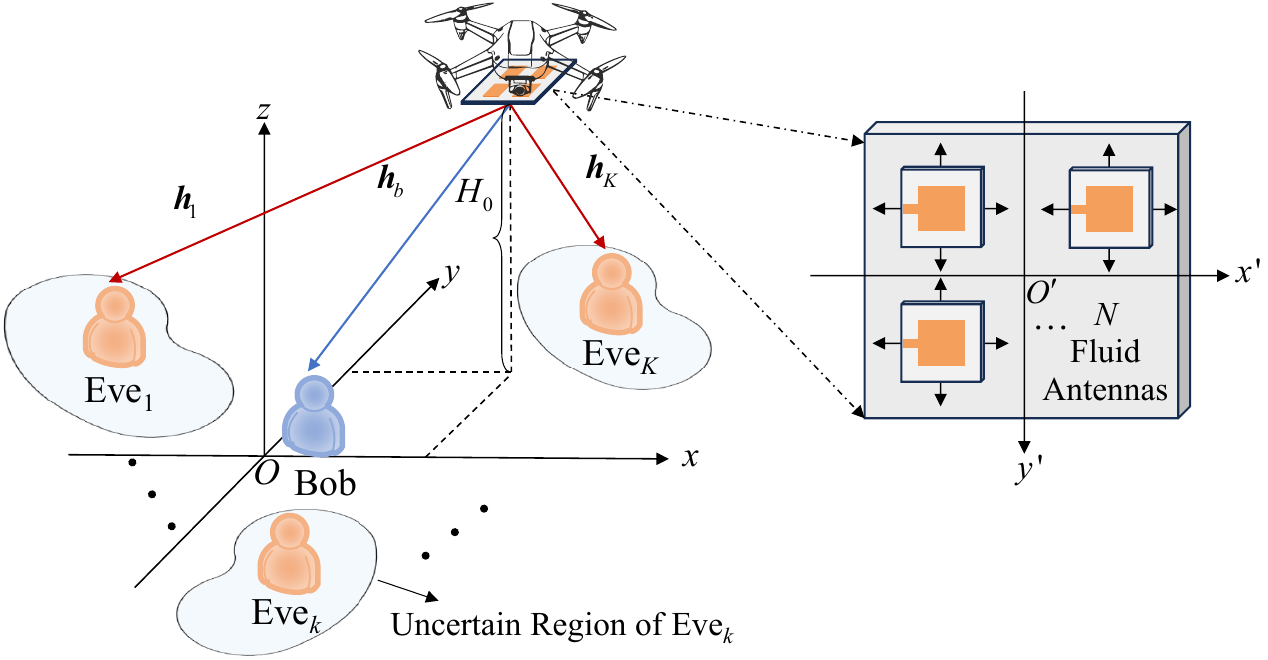}
	\caption{The FA-assisted AAV secure communication system.}
	\label{SystemModel}
\end{figure}

As shown in Fig. \ref{SystemModel}, we consider an FA-assisted secure AAV transmission system, where a rotary-wing AAV (Alice) expects to transmit confidential information to a legitimate ground user (Bob) in the presence of $ K $ eavesdroppers (Eves). Suppose that Bob and all Eves are equipped with a single FPA, while Alice is equipped with $ N $ two-dimensional (2D) FAs. Without loss of generality, independent 2D and three-dimensional (3D) Cartesian coordinate systems are established, respectively. Specifically, the 3D coordinate describes the positions of the AAV and ground receivers, where $O$ represents the origin. The 2D coordinate describes the positions of FAs w.r.t. the AAV, where $O^{\prime}$ is the origin. Due to the safety considerations such as obstacle avoidance, it is assumed that the flight altitude of the AAV is fixed as $ H_0 $\cite{3TWC21,Altitude}. The horizontal coordinate of AAV is $ \boldsymbol{q}_t=[x_t,y_t]^{\mathrm{T}} $, and Bob and all the Eves are within the plane of $ z=0 $, with their horizontal coordinates denoted by $ \boldsymbol{q}_0=[x_0,y_0 ]^{\mathrm{T}} $ and $ \boldsymbol{q}_k=[x_k,y_k ]^{\mathrm{T}},k\in\mathcal{I}_K $.

\subsection{FA Hardware Model}
\begin{figure}[t!]
	\centering
	\includegraphics[scale=0.5]{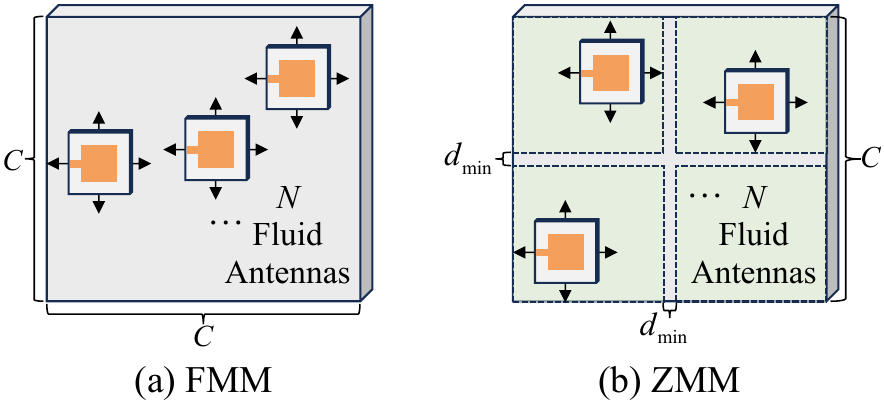}
	\vspace{-0.2cm}
	\caption{The two antenna movement modes: (a) FMM; (b) ZMM.} 
	\label{FA_mode}
\end{figure}

Denote the 2D coordinate of the $ n $-th FA by $ \boldsymbol{t}_n =[x_n,y_n]^{\mathrm{T}}, n\in\mathcal{I}_N $, and the antenna position vector (APV) of FAs can be denoted as ${\boldsymbol{t}} \triangleq {[{\boldsymbol{t}_1^{\mathrm{T}}},{\boldsymbol{t}_2^{\mathrm{T}}},... ,{\boldsymbol{t}_N^{\mathrm{T}}}]^{\mathrm{T}}}\in {\mathbb{R}^{2N\times1}}$. In this paper, we consider two movement modes for FA system, which correspond to two hardware architectures, respectively. As shown in Fig. \ref{FA_mode}(a), for the FMM, each FA can move freely within the array $\cal C$,  and the size of $\cal C$ is $ C\times C $. In order to avoid the coupling effect, a minimum allowable distance $ d_{\min} $ is required between two adjacent antennas, i.e., ${\left\| {{{\boldsymbol{t}}_n} - {{\boldsymbol{t}}_m}} \right\|_2} \ge {d_{\min }},\forall m \ne n,m\in\mathcal{I}_N$. As shown in Fig. \ref{FA_mode}(b), for the ZMM\cite{ZMM}, the $ n $-th FA is only allowed to move in a specified square region ${\cal C}_n =\{\left.[x_n,y_n]^{\mathrm{T}}\right| x_n\in[x_n^{\min},x_n^{\max} ], y_n\in[y_n^{\min}\!,y_n^{\max} ] \}$, where $x_n^{\min}$, $x_n^{\max}$, $y_n^{\min}$, and $y_n^{\max}$ are the lower and upper bounds on the $ x' $ and $ y' $ coordinates of the $ n $-th FA, respectively. The separation distance between adjacent regions is $d_{\min }$.

\subsection{Channel Model}
\begin{figure}[t!]
	\centering
	\includegraphics[scale=0.45]{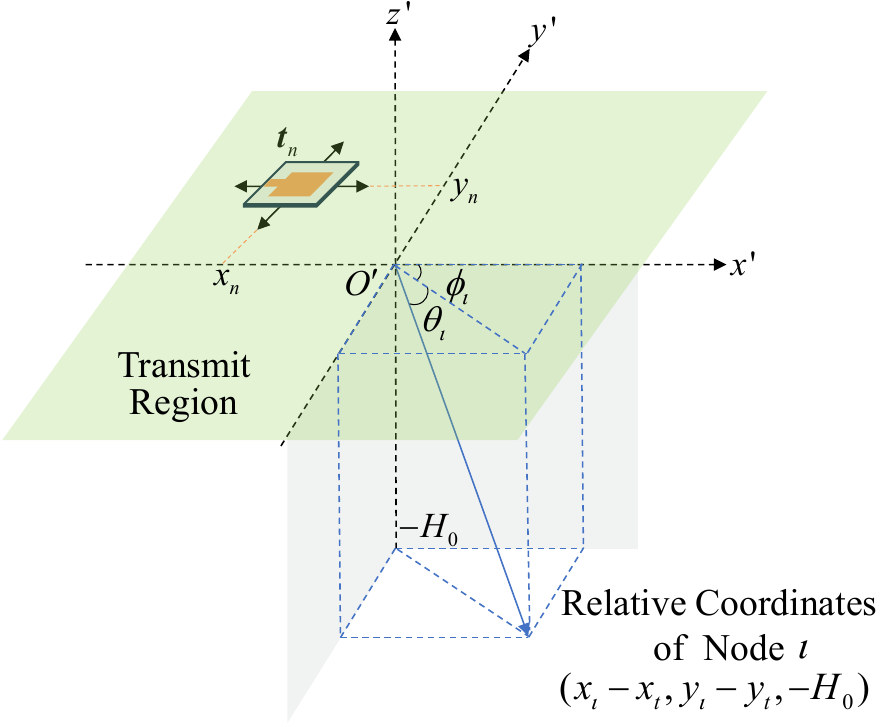}
	\caption{Illustration of 2D local coordinate and the elevation/azimuth angle corresponding to node $ \iota $.}
	\label{Coordinate}
\end{figure}

Assuming that the channels from AAV to ground receivers are dominated by LoS link\cite{LoS_TWC24,LoS_JSAC21}. Meanwhile, we focus on the scenarios in which the far-field condition is satisfied between the AAV and the ground receivers, which requires that the transmit region’s size is much smaller than the signal propagation distances. Then the channel coefficients from Alice to Bob and Eves can be uniformly denoted as
\begin{align}
	\label{channel}
	\boldsymbol{h}_{\iota}=\sqrt{\xi_0d_{\iota}^{-\alpha_{\iota}}}\boldsymbol{\alpha}^{\mathrm{H}}(\boldsymbol{t},{\boldsymbol{q}}_{t},\boldsymbol{q}_{\iota}),\iota=0,1,2,...K
\end{align}
where $\xi_0$ denotes the path loss at unit distance, $\alpha_{\iota}$ represents the path loss exponent, and $d_{\iota}=\sqrt{\left\|\boldsymbol{q}_t-\boldsymbol{q}_{\iota}\right\|_2^2+H_0^2}$ is the distance between AAV and the nodes $\iota$.
$\boldsymbol{\alpha}(\boldsymbol{t},{\boldsymbol{q}}_{t},\boldsymbol{q}_{\iota})$ is the array steering vector (ASV) of the FAs towards node $ \iota $, which depends on the relative positions of AAV and receivers. Specifically, the ASV from AAV to node $ \iota$ can be given by
\begin{align}
	\label{alpha}
	\boldsymbol{\alpha}(\boldsymbol{t},{\boldsymbol{q}}_{t},\boldsymbol{q}_{\iota})=\left[e^{j\frac{2\boldsymbol{\pi}}{\lambda}\boldsymbol{t}_1^{\mathrm{T}}\boldsymbol{\psi}_\iota},e^{j\frac{2\boldsymbol{\pi}}{\lambda}\boldsymbol{t}_2^{\mathrm{T}}\boldsymbol{\psi}_\iota},...,e^{j\frac{2\boldsymbol{\pi}}{\lambda}\boldsymbol{t}_N^{\mathrm{T}}\boldsymbol{\psi}_\iota}\right]^{\mathrm{T}}
\end{align}
where $\lambda$ is the wavelength. Herein, $\boldsymbol{\psi}_\iota$ is the unit direction vector from AAV to node $ \iota$, which is given as
\begin{align}
	\label{eq.3}
	\boldsymbol{\psi}_\iota\triangleq[\psi^x_\iota,\psi^y_\iota]^{\mathrm{T}}=\left[\cos \theta_{\iota}\cos \phi_{\iota}, \cos \theta_{\iota}\sin \phi_{\iota}\right]^{{\mathrm{T}}}.
\end{align}
As shown in Fig. \ref{Coordinate}, the trigonometric terms in \eqref{eq.3} can be determined by $\cos \theta_{\iota}=\frac{\left\|\boldsymbol{q}_{t}-\boldsymbol{q}_{\iota}\right\|_{2}}{d_{\iota}}$, $\cos \phi_{\iota}=\frac{x_{\iota}-x_{t}}{\left\|\boldsymbol{q}_{t}-\boldsymbol{q}_{\iota}\right\|_{2}}$, and $\sin \phi_{\iota}=\frac{y_{\iota}-y_{t}}{\left\|\boldsymbol{q}_{t}-\boldsymbol{q}_{\iota}\right\|_{2}}$. As a result, we can obtain $\psi_{\iota}^x=\frac{x_{\iota}-x_t}{d_{\iota}}$ and $\psi_{\iota}^y=\frac{y_{\iota}-y_t}{d_{\iota}}$.

It is assumed that the location of Bob is perfectly known by Alice, which is reasonable due to the cooperation between them. In contrast, the estimations about locations of Eves may be inaccurate and thus Alice may only know the suspicious regions where Eves possibly exist. We consider a general uncertain region model. Specifically, the uncertain region of the $k$-th Eve is represented as $\mathcal{A}_k$, whose shape can be arbitrary. Correspondingly, the uncertain position of the $k$-th Eve results in the uncertainty of ECSI $ \boldsymbol{h}_k $, which is given by
\begin{align}
	\Delta_k=\left\{\boldsymbol{h}_k\mid\boldsymbol{q}_k\in\mathcal{A}_k\right\},k\in\mathcal{I}_K.
\end{align}

\subsection{Problem Formulation}
In order to improve the achievable SR, Alice incorporates AN in the transmit signal $\boldsymbol{x}\in {\mathbb{C}^{N \times 1}}$, which can be written as $	\boldsymbol{x}=\boldsymbol{w}s+\boldsymbol{v}z$. Herein, $s\sim{\mathcal C}{\cal N}(0,1)$ and $z\sim{\mathcal C}{\cal N}(0,1)$ denote the transmit signal and the AN after normalization, respectively. ${\boldsymbol{w}} \in {\mathbb{C}^{N \times 1}}$ and  ${\boldsymbol{v}} \in {\mathbb{C}^{N \times 1}}$ represent the precoders of transmit signal and AN, respectively. The received signals at Bob and Eves can be uniformly written as
\begin{align}
	{y_{{\iota}}} = {{\boldsymbol{h}}_{{\iota}}}{\boldsymbol{w}}s + {{\boldsymbol{h}}_{{\iota}}}{\boldsymbol{v}}z + {n_{{\iota}}},\iota=0,1,2,...,K
\end{align}
where ${n_{{\iota}}}\sim{\mathcal C}{\cal N}(0,\sigma^2_{\iota})$ is the additive white Gaussian noise at node $\iota$ with average power $ \sigma^2_{\iota} $. The achievable rates of Bob and Eves can be represented, respectively, as
\begin{align}
	R_b &= \log\left(1+\frac{\left|\boldsymbol{h}_{0}\boldsymbol{w}\right|^{2}}{\left|\boldsymbol{h}_{0}\boldsymbol{v}\right|^{2}+\sigma_{0}^{2}}\right)\\
	R_k &= \log\left(1+\frac{\left|\boldsymbol{h}_{{k}}\boldsymbol{w}\right|^{2}}{\left|\boldsymbol{h}_{{k}}\boldsymbol{v}\right|^{2}+\sigma_{{k}}^{2}}\right),k\in\mathcal{I}_K.
\end{align}
From a conservative perspective, considering the most advantageous eavesdropping channel within $\Delta_{k}$ for the $k$-th Eve and the maximum eavesdropping rate among all Eves, we formulate a worst-case MSR maximization problem to provide robustness in scenarios where the ECSI exists uncertainty. To elaborate, given the signal model and the ECSI uncertainty model specified above, the worst-case MSR can be given by
\begin{align}
	\label{R_s}
	C_{s}=\left[R_{b}-\max_{k\in\mathcal{I}_K,{\boldsymbol{h}_{{k}}\in\Delta_{k}}}R_{{k}}\right]^{+}.
\end{align}

We aim to maximize $ C_s $ by jointly optimizing the precoders $ {\boldsymbol{w},\boldsymbol{v}} $, the deployment of AAV $ \boldsymbol{q}_t $, and the APV $ \boldsymbol{t} $. Accordingly, the optimization problem is formulated as
\begin{subequations}
	\begin{align}
		\label{8a}
		{\mathcal{P}}\text{0: } 
		 \max_{\boldsymbol{q}_t,\boldsymbol{w},\boldsymbol{v},\boldsymbol{t}}& \ \ C_{s}\\
		\label{8b}
		\text{s.t.} \ \ & \ \  \boldsymbol{w}^{\mathrm{H}}\boldsymbol{w}+\boldsymbol{v}^{\mathrm{H}}\boldsymbol{v}\leq P_{\mathrm{max}} \\
		\label{8c}
		& \ \ \boldsymbol{t}\in \Psi_t \\
		\label{8e}
		& \ \  x_t\in[x_t^{\min},x_t^{\max}]\\
		\label{8f}
		& \ \ y_t\in[y_t^{\min},y_t^{\max}].
	\end{align}
\end{subequations}
Herein, set $\Psi_t$ represents the feasible moving region of FAs. For the FMM and the ZMM, respectively, $\Psi_t$ is specified as
\begin{align}
	\label{constriant_FMM}
	\Psi_{t}^{\mathrm{FMM}}=&\left\{\boldsymbol{t}\mid\left\|\boldsymbol{t}_{n}-\boldsymbol{t}_{m}\right\|_{2} \geq d_{\min }, \forall m \neq n,\right. \notag\\&\ \qquad\qquad\qquad\left. \boldsymbol{t}_{n} \in \mathcal{C}, n\in\mathcal{I}_N, m\in\mathcal{I}_N\right\}\\
	\label{constriant_ZMM}
	\Psi_{t}^{\mathrm{ZMM}}=&\left\{\boldsymbol{t}\mid \boldsymbol{t}_{n} \in \mathcal{C}_n, n=1,2, \ldots, N\right\}.
\end{align}
In addition, $P_{\mathrm{max}}$ is the maximum allowable transmit power. \eqref{8e} and \eqref{8f} are the constraints on the position of AAV, where $[x_t^{\min},x_t^{\max}]$ and $[y_t^{\min},y_t^{\max}]$ denote the deployment range of AAV along $ x $-axis and $ y $-axis, respectively. Note that $\mathcal{P}0$ is intractable, and its unique challenges can be summarized as follows. First of all, the objective function is highly non-concave w.r.t. either $\boldsymbol{q}_t$, $\boldsymbol{w}$, $\boldsymbol{v}$, or $\boldsymbol{t}$. Moreover, these optimization variables are highly coupled, especially $\boldsymbol{q}_t$ and $\boldsymbol{t}$, since both variables are involved in the completed expression of the ASV. Furthermore, the uncertain regions of Eves may be irregular and meanwhile the possible values of $\boldsymbol{h}_k$ is generally an infinite set, resulting in an intractable objective function. This makes it challenging to guarantee the worst-case security performance. In what follows, we propose a robust optimization algorithm to address the aforementioned challenges, where a convex hull is constructed to deal with the uncertainty of ECSI and the worst-case MSR is maximized by invoking the AO technique.

\section{Maximization of the Worst-Case MSR}
\label{Section 3}
In this section, an AO algorithm is developed to solve $\mathcal{P}$0.  Specifically, the deployment of AAV $\boldsymbol{q}_t$, the transmit precoders ${\boldsymbol{w},\boldsymbol{v}}$, and the APV $\boldsymbol{t}$ are optimized in an alternate manner, respectively, with all the other variables being fixed. Prior to this, the uncertainties of ECSI is firstly addressed.

\subsection{Treatment of Uncertain ECSI}
Recall that our goal is to maximize the worst-case MSR for all possible eavesdropping channels within the uncertain ECSI range ${\Delta _k}$. In other words, for the $ k $-th Eve, one needs to find the most advantageous position $\boldsymbol{q}_k^{\rm opt}$ that maximizes the eavesdropping rate $ R_k $ within region  $\mathcal{A}_k$. The corresponding problem can be formulated as
\begin{align}
	\label{P0-1}
	\max_{\boldsymbol{h}_{k}\in\Delta_{k}}\log\left(1+\frac{| \boldsymbol{h}_{k}\boldsymbol{w}| ^{2}}{|\boldsymbol{h}_{k}\boldsymbol{v}|^{2}+\sigma_{k}^{2}}\right).
\end{align}
Since the logarithmic function is monotonically increasing, problem \eqref{P0-1} can be further simplified to
\begin{align}
	\label{P0-2}
	\max_{\boldsymbol{A}_k\in\Lambda_k}\frac{\boldsymbol{w}^{\mathrm{H}}\boldsymbol{A}_k\boldsymbol{w}}{\boldsymbol{v}^{\mathrm{H}}\boldsymbol{A}_k\boldsymbol{v}+\sigma_k^{\prime2}}
\end{align}
where we define $\Lambda_{k}=\left\{\boldsymbol{A}_{k}=\boldsymbol{\alpha}\left(\boldsymbol{t}, \boldsymbol{q}_{t}, \boldsymbol{q}_{k}\right) \boldsymbol{\alpha}^{\mathrm{H}}\left(\boldsymbol{t}, \boldsymbol{q}_{t}, \boldsymbol{q}_{k}\right)\mid \right. \allowbreak \left. \boldsymbol{q}_k\in\mathcal{A}_k\right\}$ and $\sigma_k^{\prime2}=\sigma_k^{2}{(\xi_0^{-1}d_{k}^{\alpha_{k}})}$. We focus on the scenarios in which the signal propagation distances from the AAV to Eves are much larger than the sizes of $\mathcal{A}_k$. In this case, $\sigma_k^{\prime2}$ can be regarded as a constant. The Dinkelbach algorithm\cite{Dinklbach} is invoked for addressing the fractional programming problem. Specifically, an auxiliary variable $\zeta$  is introduced and problem \eqref{P0-2} with given $\zeta$ can be recast as

\begin{align}
	\label{P0-3}
	\max_{\boldsymbol{A}_k\in\Lambda_k}\boldsymbol{w}^{\mathrm{H}}\boldsymbol{A}_k\boldsymbol{w}-\zeta\boldsymbol{v}^{\mathrm{H}}\boldsymbol{A}_k\boldsymbol{v}.
\end{align}
However, the objective function in \eqref{P0-3} is still non-concave. To tackle this, we construct the convex hull of $\Lambda_k$  as\cite{convex_hull}
\begin{align}
	\label{convex_hull}
	{\Omega_k} \! =\! \left\{ {{{\boldsymbol{A}}_k} \!=\! \sum\limits_{i = 1}^{{S_k}} {{\tau _{k,i}}{{\boldsymbol{A}}_{k,i}}} \left|\sum\limits_{i = 1}^{{S_k}} {{\tau _{k,i}}} \! = \! 1,{\tau _{k,i}} \ge 0\right.} \right\},\forall k
\end{align}
where $\tau_{k,i}$ is the weighted factor,  $S_k$ is the sample number, and  $\boldsymbol{A}_k$ denotes the matrix corresponding to position $ (x_k^i,y_k^i) $.

\textit{\textbf{Proposition 1}}: The following equation holds,
\begin{align}
	\label{Proposition1}
	\!\!\!\max_{\boldsymbol{A}_k\in\Lambda_k}\!\boldsymbol{w}^{\mathrm{H}} \!\boldsymbol{A}_k\boldsymbol{w}\!-\!\zeta\boldsymbol{v}^{\mathrm{H}} \!\boldsymbol{A}_k\boldsymbol{v}\!=\!\!\max_{\boldsymbol{A}_k\in\Omega_k}\!\boldsymbol{w}^{\mathrm{H}} \!\boldsymbol{A}_k\boldsymbol{w}\!-\!\zeta\boldsymbol{v}^{\mathrm{H}} \!\boldsymbol{A}_k\boldsymbol{v}.
\end{align}

\textit{Proof}: Please refer to Appendix \ref{Appendix_A}. \hfill$\blacksquare$

Armed with Proposition 1, problem \eqref{P0-3} can be converted into the following problem w.r.t. ${{\boldsymbol{\tau }}_k} = {\left[ {{\tau _{k,1}},{\tau _{k,2}},...,{\tau _{k,{S_k}}}} \right]^{\mathrm{T}}}$.
\begin{subequations}
	\label{P0-4}
	\begin{align}
		\label{tau}
		\max _{{\boldsymbol{\tau }}_k} & \ \sum_{i=1}^{S_{k}} \tau_{k, i}\left(\boldsymbol{w}^{\mathrm{H}} \boldsymbol{A}_{k, i} \boldsymbol{w}-\zeta \boldsymbol{v}^{\mathrm{H}} \boldsymbol{A}_{k, i} \boldsymbol{v}\right) \\
		\text { s.t. } & \ \sum_{i=1}^{S_{k}} \tau_{k, i}=1 \\
		& \ \tau_{k, i} \geq 0, i=1,2, \ldots, I_{k}.
	\end{align}
\end{subequations}
This problem is convex and can be solved by CVX. Denote the optimal solution of \eqref{P0-4} as ${\boldsymbol{\tau }}_k^{\rm opt}$,  and then we can obtain ${\boldsymbol{A}}_k^{{\rm{opt}}} = \sum\limits_{i = 1}^{{S_k}} {\tau _{k,i}^{{\rm{opt}}}{{\boldsymbol{A}}_{k,i}}}$ according to \eqref{convex_hull}. Subsequently, the auxiliary variable $\zeta$ is updated by
\begin{align}
	\zeta  = \frac{{{{\boldsymbol{w}}^{\mathrm{H}}}{\boldsymbol{A}}_k^{{\rm{opt}}}{\boldsymbol{w}}}}{{{{\boldsymbol{v}}^{\mathrm{H}}}{\boldsymbol{A}}_k^{{\rm{opt}}}{\boldsymbol{v}} + {\sigma ^2}}}.
\end{align}

The updated  $\zeta$ is then substituted into \eqref{tau}. Next, problem \eqref{P0-4} is solved again. This process is repeated until the relative change in  $\zeta$ between consecutive iterations is less than a predefined threshold ${\varepsilon _D} > 0$. As for how to obtain ${\boldsymbol{q}}_k^{{\rm{opt}}}$ from ${\boldsymbol{\tau}}_k^{{\rm{opt}}}$, note that the extreme points of the feasible set of problem (16) are standard basis vectors $\{\boldsymbol{e}_i\}_{i=1}^{S_k}\in\mathbb{R}^{S_k \times 1}$. Furthermore, note that the optimal solution of maximizing an affine objective function over a simplex is always one of the extreme points. As a result, the most advantageous position for the $ k $-th Eve (i.e., ${\boldsymbol{q}}_k^{{\rm{opt}}}$) can be directly obtained based on ${\boldsymbol{\tau}}_k^{{\rm{opt}}}$. For subsequent optimization stages, the optimizations of four variables are based on ${\boldsymbol{q}}_k^{{\rm{opt}}}$. In other words, the maximum eavesdropping rate of the $k$-th Eve within uncertain ECSI $\Delta_{k}$ can be denoted as $R_k^\prime =\left.R_{{k}}\right|_{{\boldsymbol{q}}_k={\boldsymbol{q}}_k^{{\rm{opt}}}}$,  and then the objective function of $\mathcal{P}0$ is transformed into ${C}_{s}^{\prime}=\left[R_{b}-\max_{{k}\in\mathcal{I}_K}R_k^\prime\right]^{+}$.

\subsection{Optimization of AAV Deployment}
\label{Section3-B}
With the given $\boldsymbol{w}$, $\boldsymbol{v}$ and $\boldsymbol{t}$, the subproblem of optimizing $\boldsymbol{q}_t$ can be formulated as
	\begin{align}
		\label{17}
		{\mathcal{P}1: } 
		\max_{\boldsymbol{q}_t}& \ \  R_{b}-\max_{{k}\in\mathcal{I}_K}R_{k}^{\prime} \\
		\text{s.t.}\ & \ \ \eqref{8e},\eqref{8f}. \notag
	\end{align}
Herein, the operator ${[ \cdot ]^ + }$ has been omitted as it does not affect the solution of the optimization problem\cite{2TCOMM25}. Note that the non-concave objective function is hard to address. In particular, the channel coefficients consist of the exponential, fractional, and radical form of $\boldsymbol{q}_t$. To tackle this, the SCA algorithm is invoked based on the constructions of affine surrogate function of $R_b$ and surrogate function of $\{\max_k{R_k^{\prime}}\}$. For such, we first rewrite $ R_b $ as
\begin{align}
	R_{b}=&\log\left(h_{w}(\boldsymbol{q}_{t})+h_{v}(\boldsymbol{q}_{t})+\sigma_{0}^{2}{h}_{d}(\boldsymbol{q}_{t})\right)\notag\\& \qquad\qquad\qquad -\log\left(h_{v}(\boldsymbol{q}_{t})+\sigma_{0}^{2}{h}_{d}(\boldsymbol{q}_{t})\right)
\end{align}
where we define $h_{w}(\boldsymbol{q}_{t}) \triangleq |\boldsymbol{\alpha}^{\mathrm{H}}(\boldsymbol{t},\boldsymbol{q}_{t},\boldsymbol{q}_0)\boldsymbol{w}|^2$, $h_{v}(\boldsymbol{q}_{t}) \triangleq |\boldsymbol{\alpha}^{\mathrm{H}}(\boldsymbol{t},\boldsymbol{q}_{t},\boldsymbol{q}_0)\boldsymbol{v}|^2$, and $h_d\triangleq\xi_{0}^{-1}d_{0}^{\alpha_{0}}$. Denote $ \boldsymbol{q}_{t}^i$ as the solution of $\mathcal{P}1$ obtained in the $ i $-th iteration of AO. Then, in the $ (i+1) $-th iteration of AO, the first-order Taylor expansion of $R_b$ w.r.t. $\boldsymbol{q}_{t}=\boldsymbol{q}_{t}^i$ can be expressed as
\begin{align}
	\bar{R}_{b}=R_{b}|_{\boldsymbol{q}_{t}=\boldsymbol{q}_{t}^{i}}+(\nabla_{\boldsymbol{q}_{t}^{i}}R_{b})^{\mathrm{T}}(\boldsymbol{q}_{t}-\boldsymbol{q}_{t}^{i}).
\end{align}
The derivation of $\nabla_{\boldsymbol{q}_{t}^{i}}R_{b}$ is given in Appendix \ref{Appendix_B}.
Until now, we have obtained an affine surrogate function $\bar{R}_b$ of $R_b$.

In a similar way, in the $ (i+1) $-th iteration of AO, an affine surrogate function of $ R_{k}^{\prime} $ can be construed as
\begin{align}
	\bar{R}_{k}=R_{k}^{\prime}|_{\boldsymbol{q}_{t}=\boldsymbol{q}_{t}^{i}}+(\nabla_{\boldsymbol{q}_{t}^{i}}R_{k}^{\prime})^{\mathrm{T}}(\boldsymbol{q}_{t}-\boldsymbol{q}_{t}^{i}), k\in\mathcal{I}_K.
\end{align}
The gradient vector $\nabla_{\boldsymbol{q}_{t}}R_{k}^{\prime}$ can be obtained by replacing $\left\{\boldsymbol{q}_{0}, \sigma_0^2\right\}$ with $\left\{\boldsymbol{q}_{k}, \sigma_k^2\right\}$ in Appendix \ref{Appendix_B}. According to the theory of convex optimization\cite{2004Convex}, $ \left\{ {\max_k {R_{k}^{\prime}}} \right\} $ retains convexity. Thus, $\mathcal{P}1$ can be converted as
\begin{align}
	\label{21}
	\mathcal{P}\text{1-1: } 
	\max_{\boldsymbol{q}_t}& \ \ \bar{R}_{b}-\max_{{k}\in\mathcal{I}_K}\bar{R}_{k}\\
	\text{s.t.} \ & \ \ \eqref{8e}, \eqref{8f}. \notag
\end{align}
$\mathcal{P}$1-1 is convex and can be solved by CVX. To address the fact that the SCA algorithm does not guarantee the convergence, the backtracking method is employed. Specifically, denote $\tilde{\boldsymbol{q}}_t^{i+1}$ as the solution of $\mathcal{P}$1-1 in the $ (i+1) $-th AO iteration, and define
$\beta\in{(0,1)}$ as the backtracking factor. Then, denote $\delta_\beta$ as the maximum value in set $\left\{\beta^j\right\}_{j=0,1,2...}$ that satisfies
\begin{align}
	C_{s}^{\prime}(\boldsymbol{q}_t^i)<C_{s}^{\prime}\left((1-\delta_\beta)\boldsymbol{q}_t^i+\delta_\beta\tilde{\boldsymbol{q}}_t^{i+1}\right).
\end{align}
Then, we adopt $\boldsymbol{q}_t^{i+1}=\left(1-\delta_\beta\right)\boldsymbol{q}_t^i+\delta_\beta\tilde{\boldsymbol{q}}_t^{i+1}$ as the final solution of $\mathcal{P}$1 in the $ (i+1) $-th AO iteration, which can ensure the MSR nondecreasing after solving $\mathcal{P}$1.

\subsection{Optimization of Transmit Precoders}
By treating $\boldsymbol{q}_t$ and $\boldsymbol{t}$ as a constant, the subproblem of optimizing $(\boldsymbol{w}, \boldsymbol{v})$ can be formulated as
	\begin{align}
		\mathcal{P}\text{2: } \max _{\boldsymbol{w},\boldsymbol{v}} \ &\ R_{b}-\max_{{k}\in\mathcal{I}_K}R_{k}^{\prime} \\
		\text{s.t. } \ &\ \eqref{8b}. \notag
	\end{align}
To proceed, we define $ {{\boldsymbol{{H}}}_0} \triangleq {\boldsymbol{h}^{\mathrm{H}}_0}{\boldsymbol{h}_0}$,  $ {{\boldsymbol{{H}}}_k} \triangleq {\boldsymbol{h}^{\mathrm{H}}_k}{\boldsymbol{h}_k}$, $ {\boldsymbol{W}} \triangleq {\boldsymbol{w}}{{\boldsymbol{w}}^{\mathrm{H}}} $, and $ {\boldsymbol{V}} \triangleq {\boldsymbol{v}}{{\boldsymbol{v}}^{\mathrm{H}}} $, which follow that $ {\boldsymbol{W}}\succcurlyeq 0$, ${\mathrm{rank}}\left( {\boldsymbol{W}} \right) = 1 $ and $ {\boldsymbol{V}}\succcurlyeq 0$, ${\mathrm{rank}}\left( {\boldsymbol{V}} \right) = 1 $. Then, the rank-one constraints are relaxed by semidefinite relaxation technique, and $\mathcal{P}$2 can be transformed to
\begin{subequations}
	\begin{align}
		\label{24a}
		\!\!\! \mathcal{P}\text{2-1: } \max _{\boldsymbol{W},\boldsymbol{V}} \ & \left\{{\log}\left( {1 + \frac{{{\mathrm{tr}}\left( {{{\boldsymbol{{H}}}_0}{\boldsymbol{W}}} \right)}}{{{\mathrm{tr}}\left( {{{\boldsymbol{{H}}}_0}{\boldsymbol{V}}} \right) + \sigma _0^2}}} \right)\right. \notag\\&\ \quad\left.- \mathop {\max_{k\in\mathcal{I}_K} } {\log}\left( {1 + \frac{{{\rm{tr}}\left( {{{\boldsymbol{{H}}}_k}{\boldsymbol{W}}} \right)}}{{{\rm{tr}}\left( {{{\boldsymbol{{H}}}_k}{\boldsymbol{V}}} \right) + \sigma _k^2}}} \right)\right\} \\
		\label{24b}
		\text{s.t. } \ & \ {\mathrm{tr}}\left(\boldsymbol{W} \right) + {\mathrm{tr}}\left( {\boldsymbol{V}} \right) \le {P_{\max }} \\
		\label{24c}
		& \ \boldsymbol{W}\succcurlyeq 0, \boldsymbol{V}\succcurlyeq 0 .
\end{align}
\end{subequations}
Note that the constraints in $\mathcal{P}$2-1 are convex, but the objective function is still non-convex and the variables are coupled. To address this, we resort Fenchel conjugate-based lemma\cite{Lemma1}.

\textit{\textbf{Lemma 1}}: Consider function $f(r)=-rx+\ln(r)+1$ for any $x>0$. Then, equality $-\ln(x)=\max_{r>0}f(r)=f(x^{-1})$ holds, with optimal solution $r^\mathrm{opt}=x^{-1}$.

By applying Lemma 1 to $  - \ln \left( {{\rm{tr}}\left( {{{\boldsymbol{{H}}}_0}{\boldsymbol{V}}} \right) + \sigma _0^2} \right) $  with $ {x_0} \triangleq {\rm{tr}}\left( {{{\boldsymbol{{H}}}_0}{\boldsymbol{V}}} \right) + \sigma _0^2 $, a equivalent expression of $R_b$ is given by eq. \eqref{eq8} at the top of the next page. Similarly, by applying Lemma 1 to $ \ln \left( {{\rm{tr}}\left( {{{\boldsymbol{{H}}}_k}\left( {{\boldsymbol{W}} + {\boldsymbol{V}}} \right)} \right) + \sigma _k^2} \right) $ with $ {x_k} \triangleq {\rm{tr}}\left( {{{\boldsymbol{{H}}}_k}\left( {{\boldsymbol{W}} + {\boldsymbol{V}}} \right)} \right) + \sigma _k^2 $, a equivalent expression of $R_k^\prime$ is given by eq. \eqref{eq9} at the top of the next page.
\begin{figure*}[!t]
	\normalsize
	\setcounter{MYtempeqncnt}{\value{equation}}
	\setcounter{equation}{26}	
	\begin{equation}
		R_{b} \ln 2 = \max _{r_{0}>0}\left[\ln \left(\operatorname{tr}\left(\boldsymbol{H}_{0} \boldsymbol{W}\right)+\operatorname{tr}\left(\boldsymbol{H}_{0} \boldsymbol{V}\right)+\sigma_{0}^{2}\right)-r_{0}\left(\operatorname{tr}\left(\boldsymbol{H}_{0} \boldsymbol{V}\right)+\sigma_{0}^{2}\right)+\ln r_{0}+1\right]\triangleq \max _{r_{0}>0} f_{0}\left(\boldsymbol{W}, \boldsymbol{V}, r_{0}\right).
		\label{eq8}
	\end{equation}
    \begin{equation}
	    R_{k}^{\prime} \ln 2 =\min _{r_{k}>0}\left[r_{k}\left(\operatorname{tr}\left(\boldsymbol{H}_{k} \boldsymbol{W}\right)+\operatorname{tr}\left(\boldsymbol{H}_{k} \boldsymbol{V}\right)+\sigma_{k}^{2}\right)-\ln r_{k}-1-\ln \left(\operatorname{tr}\left(\boldsymbol{H}_{k} \boldsymbol{V}\right)+\sigma_{k}^{2}\right)\right]\triangleq \min _{r_{k}>0} f_{k}\left(\boldsymbol{W}, \boldsymbol{V}, r_{k}\right).
	    \label{eq9}
    \end{equation}  
	\hrulefill
	\setcounter{equation}{\value{MYtempeqncnt}}
\end{figure*}
Therefore, $\mathcal{P}$2-1 can be equivalently expressed as
\setcounter{equation}{28} 
\begin{subequations}
	\begin{align}
		\label{eq27a}
		\!\!\mathcal{P}\text{2-2: } \!\!\! \max_{{\boldsymbol{W}},{\boldsymbol{V}},{r_0},{r_k}}\! &\Big( {{f_0}\left( {{\boldsymbol{W}},{\boldsymbol{V}},{r_0}} \right)\!- \!\max_{k\in\mathcal{I}_K} {f_k}\left( {{\boldsymbol{W}},{\boldsymbol{V}},{r_k}} \right)} \Big) \\
		\label{eq27b}
		\text{s.t. } \ \ &\ {\mathrm{tr}}\left(\boldsymbol{W} \right) + {\mathrm{tr}}\left( {\boldsymbol{V}} \right) \le {P_{\max }} \\
		\label{eq27c}
		&\ \boldsymbol{W}\succcurlyeq 0, \boldsymbol{V}\succcurlyeq 0 \\
		\label{eq27d}
		&\ r_0>0 , r_k>0, k\in\mathcal{I}_K.
	\end{align}
\end{subequations}

$\mathcal{P}$2-2 is concave w.r.t either $({\boldsymbol{W}},{\boldsymbol{V}})$ or $ ({r_0},{r_k}) $ by fixing the other variables. This motivates us to utilize the AO algorithm. According to Lemma 1, the optimal solutions of $ ({r_0},{r_k}) $ with fixed $({\boldsymbol{W}},{\boldsymbol{V}})$ are 
\begin{align}
	\label{r_b}
	 &\qquad \ r_0^\mathrm{opt} = {\left( {{\rm{tr}}\left( {{{\boldsymbol{{H}}}_0}{\boldsymbol{V}}} \right) + \sigma _b^2} \right)^{ - 1}}\\
	\label{r_k}
	&r_k^\mathrm{opt} = {\left( {{\rm{tr}}\left( {{{\boldsymbol{{H}}}_k}{\boldsymbol{W}}} \right) + {\rm{tr}}\left( {{{\boldsymbol{{H}}}_k}{\boldsymbol{V}}} \right) + \sigma _k^2} \right)^{-1}}.
\end{align}
As for the optimal solutions of $({\boldsymbol{W}},{\boldsymbol{V}})$ with given $ ({r_0^\mathrm{opt}},{r_k^\mathrm{opt}}) $, the subproblem is concave and can be solved efficiently by CVX. Finally, the Gaussian randomization method is utilized to recover the solution for $\mathcal{P}$2\cite{Lemma1}.

\subsection{Optimization of FA positions}
For the FMM and the ZMM, respectively, distinct design schemes of FA positions are proposed in what follows.

1) \textit{Optimizing $ \boldsymbol{t} $ under FMM}

In the FMM, to avoid the coupling effects, the position of each FA $ \{\boldsymbol{t}_n\}_{n=1}^N $ is alternately optimized.
Hence, the subproblem of optimizing $\boldsymbol{t}_n$ can be written as
\begin{subequations}
	\begin{align}
		\label{eq30}
		\mathcal{P}\text{3-1: } 
		\max_{\boldsymbol{t}_n}& \ \ R_{b}-\max_{{k}\in\mathcal{I}_K}R_{k}^{\prime}\\
		\label{eq30b}
		\text{s.t.}& \ \ \left\|\boldsymbol{t}_{n}-\boldsymbol{t}_{m}\right\|_{2} \geq d_{\min }, \forall m \neq n \\
		\label{eq30c}
		& \ \  \boldsymbol{t}_{n} \in \mathcal{C}, n\in\mathcal{I}_N.
	\end{align}
\end{subequations}
Note that the objective function \eqref{eq30} and the constraint \eqref{eq30b} are non-concave w.r.t. $\boldsymbol{t}_n$. To tackle this problem, we construct concave surrogate functions of \eqref{eq30} and \eqref{eq30b}. To this end, we first rewrite the $ R_b $ and $R_{k}^{\prime} $ as
\begin{align}
	\label{31}
	{R_b} =& \ {{{\log }}\left( {|{{\boldsymbol{\alpha }}^{\mathrm{H}}}({\boldsymbol{t}},\boldsymbol{q}_t,\boldsymbol{q}_0){\boldsymbol{w}}{|^2} + |{{\boldsymbol{\alpha }}^{\mathrm{H}}}({\boldsymbol{t}},\boldsymbol{q}_t,\boldsymbol{q}_0){\boldsymbol{v}}{|^2} + \sigma_0^{\prime2}} \right)} \notag\\&\qquad- {{{\log }}\left( {|{{\boldsymbol{\alpha }}^{\mathrm{H}}}({\boldsymbol{t}},\boldsymbol{q}_t,\boldsymbol{q}_0){\boldsymbol{v}}{|^2} + \sigma_0^{\prime2}} \right)} \notag\\ \triangleq& {{I_{b,1}}}-{{I_{b,2}}} \\
	\label{32}
	{R_{k}^{\prime}} =& \ {{{\log }}\left( {|{{\boldsymbol{\alpha }}^{\mathrm{H}}}({\boldsymbol{t}},\boldsymbol{q}_t,\boldsymbol{q}_k^\mathrm{opt}){\boldsymbol{w}}{|^2} + |{{\boldsymbol{\alpha }}^{\mathrm{H}}}({\boldsymbol{t}},\boldsymbol{q}_t,\boldsymbol{q}_k^\mathrm{opt}){\boldsymbol{v}}{|^2} + \sigma_k^{\prime2}} \right)} \notag\\&\qquad- {{{\log }}\left( {|{{\boldsymbol{\alpha }}^{\mathrm{H}}}({\boldsymbol{t}},\boldsymbol{q}_t,\boldsymbol{q}_k^\mathrm{opt}){\boldsymbol{v}}{|^2} + \sigma_k^{\prime2}} \right)} \notag\\ \triangleq& {{I_{k,1}}}-{{I_{k,2}}}
\end{align}
where $\sigma_0^{\prime2} = \sigma _0^2{{\xi _0^{-1}}d_0^{ {\alpha _0}}}$ and $\sigma_k^{\prime2}=\sigma_k^{2}{\xi_0^{-1}d_{k}^{\alpha_{k}}}$. Our goal is to obtain a concave surrogate function of $ R_b $ and a convex surrogate function of $ R_{k}^{\prime} $. For such, in what follows,  concave surrogate functions of $ I_{b,1} $ and $ I_{k,2} $ as well as  affine surrogate functions of $ I_{b,2} $ and $ I_{k,1} $ are derived.

To begin with, we derive the concave lower bound of $ I_{b,1} $. For such, define $\alpha_{b,n}$ and $ w_n $ as the $ n $-th element of $\boldsymbol{\alpha}(\boldsymbol{t},\boldsymbol{q}_{t},\boldsymbol{q}_{0})$ and ${\boldsymbol{w}}$, respectively. Then, $| \boldsymbol{\alpha}^{\mathrm{H}}(\boldsymbol{t},\boldsymbol{q}_{t},\boldsymbol{q}_{0})\boldsymbol{w}| ^2$ can be rewritten as
\begin{align}
	\label{h_bw}
	\left| \boldsymbol{\alpha}^{\mathrm{H}}(\boldsymbol{t},\boldsymbol{q}_{t},\boldsymbol{q}_{0})\boldsymbol{w}\right| ^2  &=2\operatorname{Re}\left\{\alpha_{b,n}^{*}w_{n}\Xi_{b,w}^{*}\right\}+\left|w_{n}\right|^{2}+\left|\Xi_{b,w}\right|^{2} \notag\\
	&\triangleq h_{b,w}(\boldsymbol{t}_{n})
\end{align}
where $\Xi_{b,w}\triangleq\sum\nolimits_{m \ne n}^N {{\alpha^*_{b,m}}{w_m}} $ is a constant complex value and is independent of $\boldsymbol{t}_{n}$. Similarly, define $v_n$ as the $n$-th element of $\boldsymbol{v}$, and $| \boldsymbol{\alpha}^{\mathrm{H}}(\boldsymbol{t},\boldsymbol{q}_{t},\boldsymbol{q}_{0})\boldsymbol{v}| ^2$ can be rewritten as
\begin{align}
	\label{h_bv}
	\left| \boldsymbol{\alpha}^{\mathrm{H}}(\boldsymbol{t},\boldsymbol{q}_{t},\boldsymbol{q}_{0})\boldsymbol{v}\right| ^2 &=2\operatorname{Re}\left\{\alpha_{b,n}^{*}v_{n}\Xi_{b,v}^{*}\right\}+\left|v_{n}\right|^{2}+\left|\Xi_{b,v}\right|^{2} \notag \\ &\triangleq h_{b,v}(\boldsymbol{t}_{n})
\end{align}
where $\Xi_{b,v}\triangleq\sum\nolimits_{m \ne n}^N {{\alpha^*_{b,m}}{v_m}} $ is also a constant. Combining \eqref{alpha}, \eqref{h_bw}, and \eqref{h_bv}, we have
\begin{align}
	&h_{b,w}\left(\boldsymbol{t}_{n}\right)+h_{b,v}\left(\boldsymbol{t}_{n}\right) \notag\\
	&\qquad=2|\Xi_{b,wv}|\cos\left(\rho_{b}x_{n}+\eta_{b}y_{n}-\angle\Xi_{b,wv}\right)+\Phi_{b} \notag\\
	& \qquad\triangleq2g_{b}\left(\boldsymbol{t}_{n}\right)+\Phi_{b}
\end{align}
where ${\Xi _{b,wv}} \triangleq {w_n}{\Xi^* _{b,w}} + {v_n}{\Xi^* _{b,v}} \triangleq \left| {{\Xi _{b,wv}}} \right|{e^{j\angle {\Xi _{b,wv}}}}$, $\rho_b \triangleq \frac{2\pi}{\lambda} \psi_0^x$, $\eta_{b}\triangleq \frac{2\pi}{\lambda} \psi_0^y$, $\Phi_{b} \triangleq {\left| {{w_n}} \right|^2} + {\left| {{\Xi _{b,w}}} \right|^2} + {\left| {{v_n}} \right|^2} + {\left| {{\Xi _{b,v}}} \right|^2}$. According to the composition rule\cite{2004Convex}, constructing the concave surrogate function of $ I_{b,1} $ can be simplified to constructing the concave surrogate function of $g_{b}\left(\boldsymbol{t}_{n}\right)$. For such, the following lemma is proposed.

\textit{\textbf{Lemma 2}}: A global concave lower bound for $g_{b}\left(\boldsymbol{t}_{n}\right)$ can be constructed as
\begin{align}
	g_{b}\left(\boldsymbol{t}_n\right)\geq&-\frac{\gamma_{b,n}}{2}\boldsymbol{t}_n^{\mathrm{T}}\boldsymbol{t}_n+\left(\left(\nabla_{\boldsymbol{t}_n^i}g_{b}\right)^{\mathrm{T}}+\gamma_{b,n}\boldsymbol{t}_n^{i{\mathrm{T}}}\right)\boldsymbol{t}_n \notag\\ &\qquad+g_{b}\left(\boldsymbol{t}_n^i\right)-\left(\nabla_{\boldsymbol{t}_n^i}g_{b}\right)^{\mathrm{T}}\boldsymbol{t}_n^i-\frac{\gamma_{b,n}}{2}\boldsymbol{t}_n^{i{\mathrm{T}}}\boldsymbol{t}_n^i \notag\\ \triangleq& \ \tilde{g}_{b}\left(\boldsymbol{t}_n\right)
\end{align}
where $\boldsymbol{t}_n^{i}$ denotes the solution of $\mathcal{P}$3-1 obtained in the $ i $-th iteration of the AO
. The positive real number ${\gamma _{b,n}} = \left| {{\Xi _{b,wv}}} \right|\left( {\rho_b^2 + \eta_b^2} \right)$ satisfies ${\gamma _{b,n}}{{\boldsymbol{I}}_2}\succcurlyeq\nabla _{{{\boldsymbol{t}}_n}}^2{g_{b}}$.

\textit{Proof}: Please refer to Appendix \ref{Appendix_C}. \hfill$\blacksquare$

Armed with Lemma 2, the concave surrogate function that provides a global lower bound for  $ I_{b,1} $ can be constructed as
\begin{align}
	\tilde{I}_{b,1}\triangleq\log\left(2\tilde{g}_{b}(\boldsymbol{t}_n)+\Phi_{b}+\sigma_0^{\prime2}\right).
\end{align}

Next, to construct the affine surrogate function of $ I_{b,2} $, we firstly derive the gradient of $ I_{b,2} $ as follows.
\begin{align}
	\nabla_{\boldsymbol{t}_n}I_{b,2}=\frac{1}{\ln2}{\left[\frac{\nabla_{\boldsymbol{t}_n}h_{b,v}}{h_{b,v}(\boldsymbol{t}_n)+\sigma_0^{\prime2}}\right]}
\end{align}
where $\nabla_{\boldsymbol{t}_n}h_{b,v}$ can be derived by following a similar derivation process as that in Appendix \ref{Appendix_C}. Then, the surrogate function of $I_{b,2}$ can be given by
\begin{align}
	\tilde{I}_{b,2}\triangleq\log\left(h_{b,v}\left(\boldsymbol{t}_n^i\right)+\sigma_0^{\prime2}\right)+\left(\nabla_{\boldsymbol{t}_n^i}I_{b,2}\right)^{\mathrm{T}}\left(\boldsymbol{t}_n-\boldsymbol{t}_n^i\right).
\end{align}

Until now, a concave surrogate function of $ R_b $ has been obtained, denoted as
$\tilde{R}_b=\tilde{I}_{b,1}-\tilde{I}_{b,2}$. Note that the affine approximation function ($\tilde{I}_{k,1}$) for $I_{k,1}$ and the concave lower bound ($\tilde{I}_{k,2}$) for $ I_{k,2} $ can be derived by adopting the derivation methodology used for $ \tilde{I}_{b,2} $ and $ \tilde{I}_{b,1} $, respectively. Consequently, we obtain a convex surrogate function for $ R_{k}^{\prime} $, denoted as
$\tilde{R}_k=\tilde{I}_{k,1}-\tilde{I}_{k,2}$. While the specific derivations for $\tilde{I}_{k,1}$ and $\tilde{I}_{k,2}$ are omitted here for brevity, the involved Hessian matrix and gradient vector can be obtained by replacing $ \boldsymbol{q}_0 $ with $ \boldsymbol{q}_k $ in Appendix \ref{Appendix_C}. Finally, we construct the concave approximation function of objective function \eqref{eq30} in $\mathcal{P}$3-1, denoted as $\tilde{R}_b-\max_{k\in\mathcal{I}_K}\{\tilde{R}_k\}$. However, the constraint \eqref{eq30b} is still non-convex. To address this, in each iteration, ${f_t}({{\boldsymbol{t}}_n}) \triangleq \left\| {{{\boldsymbol{t}}_n} - {{\boldsymbol{t}}_m}} \right\|_2$ is replaced by its lower bound, which is constructed by first-order approximation as follows
\begin{align}
	f_t(\boldsymbol{t}_n)&\geq f_t(\boldsymbol{t}_n^i)+\left(\nabla_{\boldsymbol{t}_n^i}f_t(\boldsymbol{t}_n)\right)^{\mathrm{T}}\left(\boldsymbol{t}_n-\boldsymbol{t}_n^i\right)\notag\\&=\frac{\left(\boldsymbol{t}_n^i-\boldsymbol{t}_m\right)^{\mathrm{T}}\left(\boldsymbol{t}_n-\boldsymbol{t}_m\right)}{\left\|\boldsymbol{t}_n^i-\boldsymbol{t}_m\right\|_2}.
\end{align}

Based on the derivations above, for the $ (i+1) $-th iteration of the AO algorithm, $\mathcal{P}$3-1 can be converted as
\begin{subequations}
	\begin{align}
		\label{P3-1-A}
		\!\!\!\!\mathcal{P}\text{3-1-A: }  \max_{\boldsymbol{t}_{n}}&\ \tilde{R}_{b}-\max_{k\in\mathcal{I}_K}\tilde{R}_{k} \\
		\mathrm{s.t.} & \  \frac{\left(\boldsymbol{t}_n^i-\boldsymbol{t}_m\right)^{\mathrm{T}} \!\! \left(\boldsymbol{t}_n-\boldsymbol{t}_m\right)}{\left\|\boldsymbol{t}_n^i-\boldsymbol{t}_m\right\|_2}\geq d_{\min},\forall m\neq n \\
		& \ \boldsymbol{t}_{n} \in \mathcal{C}, n\in\mathcal{I}_N.
	\end{align}
\end{subequations}
$\mathcal{P}$3-1-A is convex and can be efficiently solved by CVX. Furthermore, the backtracking method is utilized to guarantee convergence, which has been described in Section \ref{Section3-B}, so the details are omitted for brevity.

2) \textit{Optimizing $ \boldsymbol{t} $ under ZMM}

In the ZMM, since the distances between the sub-regions $\mathcal{C}_n$ satisfy the minimum allowable distance $d_{\min}$, the subproblem of optimizing FA positions can be written as
\begin{subequations}
	\begin{align}
		\label{P3-2}
		\mathcal{P}\text{3-2: } 
		\max_{\boldsymbol{t}}& \ \  R_{b}-\max_{{k}\in\mathcal{I}_K}R_{k}^{\prime} \\
		\text{s.t.}& \ \ \boldsymbol{t}_{n}\in\mathcal{C}_n, n\in\mathcal{I}_N.
	\end{align}
\end{subequations}
The ADMM algorithm is introduced to address $\mathcal{P}$3-2. Specifically, by introducing auxiliary variable ${\boldsymbol{s}} = {\left[ {{s_1},s_2, \ldots ,{s_K}} \right]^{\mathrm{T}}}$, $\mathcal{P}$3-2 can be equivalently written as
\begin{subequations}
	\begin{align}
		\label{P3-2-A}
		\mathcal{P}\text{3-2-A:}
		\min_{\boldsymbol{t},\boldsymbol{s}} &\ \ -R_b+\max_{k\in\mathcal{I}_K}{s_k} \\
		\label{43b}
		\mathrm{s.t.} & \ \ s_k=R_{k}^{\prime}, k\in\mathcal{I}_K \\
		\label{43c}
		& \ \ \boldsymbol{t}_{n}\in\mathcal{C}_n, n\in\mathcal{I}_N.
	\end{align}
\end{subequations}
For $\mathcal{P}$3-2-A, its augmented Lagrangian function is given by
\begin{align}
	\mathcal{L}_\rho(\boldsymbol{t},\boldsymbol{s},\boldsymbol{\mu})&=-R_b+\max_{k\in\mathcal{I}_K}{s_k}\notag \\ &+\sum_{k=1}^K\left[\mu_k\left(s_k-R_{k}^{\prime}\right)+\frac{\rho}{2}\left(s_k-R_{k}^{\prime}\right)^2\right]
\end{align}
where ${\boldsymbol{\mu }} = {\left[ {{\mu _1},{\mu_2},...,{\mu _K}} \right]^{\mathrm{T}}}$ denotes the Lagrangian multiplier associated with the equality constraint \eqref{43b}, and $\rho>0$ is the quadratic penalty parameter to improve the stability and convergence of algorithm. Then, the corresponding Lagrangian dual problem can be written as
\begin{subequations}
	\begin{align}
		\label{P3-2-B}
		\mathcal{P}\text{3-2-B:}
		\max_{\boldsymbol{\mu}}\min_{\boldsymbol{t},\boldsymbol{s}} &\ \ \mathcal{L}_\rho(\boldsymbol{t},\boldsymbol{s},\boldsymbol{\mu}) \\
		\label{45b}
		\mathrm{s.t.} & \ \ \boldsymbol{t}_{n}\in\mathcal{C}_n, n\in\mathcal{I}_N.
	\end{align}
\end{subequations}
Let $ (\boldsymbol{t}^0,\boldsymbol{s}^0,\boldsymbol{\mu}^0) $ denote the initial primal-dual variables. In the $ (i_A+1) $-th iteration of ADMM algorithm, the standard ADMM consists of the following iterative procedures
\begin{empheq}[left=\empheqlbrace]{align}
	\boldsymbol{t}^{i_A+1} &= \arg\min_{\boldsymbol{t}_n \in \mathcal{C}_n} \mathcal{L}_\rho(\boldsymbol{t}, \boldsymbol{s}^{i_A}, \boldsymbol{\mu}^{i_A}) \label{eq:step1} \\
	\boldsymbol{s}^{i_A+1} &= \arg\min_{\boldsymbol{s} \geq 0} \mathcal{L}_\rho(\boldsymbol{t}^{i_A+1}, \boldsymbol{s}, \boldsymbol{\mu}^{i_A}) \label{eq:step2} \\
	\mu_k^{i_A+1} &= \mu_k^{i_A} + \rho\left(s_k^{i_A+1} - R_{k}^{\prime}(\boldsymbol{t}^{i_A+1})\right), k\in\mathcal{I}_K. \label{eq:step3}
\end{empheq}

Note that the minimization problem \eqref{eq:step2} is convex, so its optimal solution can be obtained by CVX. Next, we deal with the minimization problem \eqref{eq:step1}. By ignoring the constant term ${\max _k}{s_k}$, the subproblem w.r.t. $\boldsymbol{t}$ can be rewritten as 
\begin{subequations}
	\begin{align}
		\label{P3-2-C}
		\!\!\!\mathcal{P}\text{3-2-C:}
		\min_{\boldsymbol{t}} &\ 	- \!\! R_b \!+\! \sum_{k=1}^K \!\!\left[\mu_k\left(s_k \!- \! R_{k}^{\prime}\right) \!+\! \frac{\rho}{2}\left(s_k \!-\! R_{k}^{\prime}\right)^2\right] \\
		\mathrm{s.t.} & \ \boldsymbol{t}_{n}\in\mathcal{C}_n, n\in\mathcal{I}_N.
	\end{align}
\end{subequations}
To address problem $\mathcal{P}$3-2-C, a concave surrogate function of ${R_b} \triangleq {I_{b,1}} - {I_{b,2}}$ is constructed. To this end, the following lemma is proposed.

\textit{\textbf{Lemma 3}}: The concave surrogate function of $ I_{b,1} $ is given by
\begin{align}
	\label{I_b,1}
	\hat{I}_{b,1}\triangleq\log\left(\boldsymbol{L}_w^{\mathrm{T}}\boldsymbol{t}+c_w+\boldsymbol{L}_v^{\mathrm{T}}\boldsymbol{t}+c_v+\sigma_0^{\prime2}\right)
\end{align}
where $ \boldsymbol{L}_w \in\mathbb{R}^{2N \times1} $, $\boldsymbol{L}_v \in\mathbb{R}^{2N \times1} $, $c_w $, and $c_v $ are given by
\begin{align}
	&L_w[\hat{n}] = -2l(\hat{n}) \sum_{m=1}^N | w_n w_m | \sin  \left( f_w (\boldsymbol{t}_n^{i_A},\boldsymbol{t}_m^{i_A}) \right)\\
	&L_v[\hat{n}] = -2l(\hat{n}) \sum_{m=1}^N | v_n v_m |  \sin  \left( f_v (\boldsymbol{t}_n^{i_A},\boldsymbol{t}_m^{i_A}) \right)\\
	&c_{w} \! =\!\!\sum_{n=1}^N \!\sum_{m=1}^N \!\!|w_nw_m|\!\!\left[\cos\!\left(f_w(\boldsymbol{t}_n^{i_A},\boldsymbol{t}_m^{i_A})\!\right) 	\!\!+\! \rho_b\sin\!\left(f_w(\boldsymbol{t}_n^{i_A},\boldsymbol{t}_m^{i_A})\!\right)\right.\notag\\& \left. \quad \times \left(x_n^{i_A}\!-\! x_m^{i_A}\right)
	\!+\!\eta_b\sin\left(f_w(\boldsymbol{t}_n^{i_A},\boldsymbol{t}_m^{i_A})\right)\left(y_n^{i_A}-y_m^{i_A}\right)\right] \\
	&c_{v} \! =\!\!\sum_{n=1}^N \!\sum_{m=1}^N \!|v_nv_m|\!\!\left[\cos\!\left(f_v(\boldsymbol{t}_n^{i_A},\boldsymbol{t}_m^{i_A})\!\right) 	\!+\! \rho_b\sin\left(f_v(\boldsymbol{t}_n^{i_A},\boldsymbol{t}_m^{i_A})\!\right)\right.\notag\\& \left. \quad \times \left(x_n^{i_A}\!-\! x_m^{i_A}\right)
	\!+\!\eta_b\sin\left(f_v(\boldsymbol{t}_n^{i_A},\boldsymbol{t}_m^{i_A})\right)\left(y_n^{i_A}-y_m^{i_A}\right)\right]
\end{align}
with $l(\hat{n}) \!=\! \rho_b$ for $\hat{n} \!=\! 2n\!-\!1$ and $l(\hat{n})\!=\! \eta_b$ for $\hat{n} \!=\! 2n$. $\boldsymbol{t}^{i_A} \triangleq \allowbreak\\ \left[{\boldsymbol{t}_1^{i_A{\mathrm{T}}}}\!\!,{\boldsymbol{t}_2^{i_A{\mathrm{T}}}}\!\!,...,{\boldsymbol{t}_N^{i_A{\mathrm{T}}}}\right]^{\mathrm{T}} \!\!\!\!=\![x_1^{i_A}\!,y_1^{i_A}\!,x_2^{i_A}\!,y_2^{i_A}\!,...,x_N^{i_A}\!,y_N^{i_A}]^{\mathrm{T}}$ denotes the solution of $ \boldsymbol{t} $ obtained in the $ i_A $-th iteration of the ADMM. 

\textit{Proof}: Based on the expansion derived in \eqref{31}, $|{{\boldsymbol{\alpha }}^{\mathrm{H}}}({\boldsymbol{t}},\boldsymbol{q}_t,\boldsymbol{q}_0){\boldsymbol{w}}{|^2}$ and $|{{\boldsymbol{\alpha }}^{\mathrm{H}}}({\boldsymbol{t}},\boldsymbol{q}_t,\boldsymbol{q}_0){\boldsymbol{v}}{|^2}$ can be rewritten as
\begin{align}
	\label{eq55}
	&|\boldsymbol{\alpha}^{\mathrm{H}}(\boldsymbol{t},\boldsymbol{q}_t,\boldsymbol{q}_0)\boldsymbol{w}|^2 = \sum _{n = 1}^N \sum _{m = 1}^N  \left| {{w_n}{w_m}} \right|\times\notag\\ &\qquad\cos \underbrace{\left[\rho_{b}\left( x_{n} \!-\! x_{m}\right) \!+\!\eta_{b}\left( y_{n}\!-\! y_{m}\right) \!-\! \left( \angle w_{n}\!-\!\angle w_{m}\right) \right]  }_{f_{w}\left( \boldsymbol{t}_n,\boldsymbol{t}_m\right) }\\
	\label{eq56}
	&|\boldsymbol{\alpha}^{\mathrm{H}}(\boldsymbol{t},\boldsymbol{q}_t,\boldsymbol{q}_0)\boldsymbol{v}|^2 = \sum _{n = 1}^N \sum _{m = 1}^N  \left| {{v_n}{v_m}} \right|\times\notag\\ &\qquad\cos \underbrace{\left[\rho_{b}\left( x_{n} \!-\! x_{m}\right) \!+\!\eta_{b}\left( y_{n}\!-\! y_{m}\right) \!-\! \left( \angle v_{n}\!-\!\angle v_{m}\right) \right]  }_{f_{v}\left( \boldsymbol{t}_n,\boldsymbol{t}_m\right) }
\end{align}
where $|{w_n}|$ and $ \angle {w_n} $ represent the amplitude and the phase of $w_n$, respectively, whereas $|{v_n}|$ and $ \angle {v_n} $ is the amplitude and the phase of $v_n$, respectively.
Then, the proof can be completed by the first-order Taylor expansions of $\cos\left({f_{w}\left( \boldsymbol{t}_n,\boldsymbol{t}_m\right) }\right)$ and $\cos\left({f_{v}\left( \boldsymbol{t}_n,\boldsymbol{t}_m\right) }\right)$ w.r.t $\left( \boldsymbol{t}_n,\boldsymbol{t}_m\right)$ at $(\boldsymbol{t}_n^{i_A},\boldsymbol{t}_m^{i_A})$, respectively. \hfill$\blacksquare$

Next, we derive an affine local approximation of $ I_{b,2} $. 
For the $ (i_A+1) $-th iteration of ADMM, the local affine approximation of $ I_{b,2} $ at $\boldsymbol{t}=\boldsymbol{t}^{i_A}$ can be constructed as
\begin{align}
	\label{I_b,2}
	\!{\hat I_{b,2}} \!\triangleq \!{\log}\left( {|{{\boldsymbol{\alpha }}^{\mathrm{H}}}({{\boldsymbol{t}}^{i_A}},\boldsymbol{q}_t,\boldsymbol{q}_0){\boldsymbol{v}}{|^2} \!+\! \sigma _0^{\prime2}} \right) \!+\! {\left( {{\nabla _{{{\boldsymbol{t}}^{i_{A}}}}}{I_{b,2}}} \right)^{\mathrm{T}}}\!\left({{\boldsymbol{t}} \!-\! {{\boldsymbol{t}}^{i_A}}}\right).
\end{align}
Herein, ${{\nabla _{{{\boldsymbol{t}}}}}{I_{b,2}}}$ denotes the gradient of $ I_{b,2} $, which is given by
\begin{align}
	{\nabla_{\boldsymbol{t}}}{I_{b,2}}=\frac1{\ln2}{\left[\frac{\boldsymbol{\mathit{\Gamma}}_{b,v}^{(x)}\otimes \boldsymbol{e}_1+\boldsymbol{\mathit{\Gamma}}_{b,v}^{(y)}\otimes \boldsymbol{e}_2}{\left| \boldsymbol{\alpha}^{\mathrm{H}}(\boldsymbol{t},\boldsymbol{q}_{t},\boldsymbol{q}_{0})\boldsymbol{v}\right| ^2+\sigma_0^{\prime2}}\right]}.
\end{align}
We define $\{\boldsymbol{e}_i\}_{i=1}^2\in\mathbb{R}^{2 \times 1}$, $\boldsymbol{\mathit{\Gamma}}_{b,v}^{(x)} \triangleq-\boldsymbol{A}_{b,R}^{(x)}\big(2\boldsymbol{R}_v\boldsymbol{\alpha}_{b,R}+2\boldsymbol{P}_v\boldsymbol{\alpha}_{b,I}\big)+\boldsymbol{A}_{b,I}^{(x)}\big(2\boldsymbol{R}_v\boldsymbol{\alpha}_{b,I}+2\boldsymbol{P}_v^{\mathrm{T}}\boldsymbol{\alpha}_{b,R}\big)$,  $\boldsymbol{\mathit{\Gamma}}_{b,v}^{(y)} \triangleq-\boldsymbol{A}_{b,R}^{(y)}\big(2\boldsymbol{R}_v\boldsymbol{\alpha}_{b,R}+2\boldsymbol{P}_v\boldsymbol{\alpha}_{b,I}\big)+\boldsymbol{A}_{b,I}^{(y)}\big(2\boldsymbol{R}_v\boldsymbol{\alpha}_{b,I}+2\boldsymbol{P}_v^{\mathrm{T}}\boldsymbol{\alpha}_{b,R}\big)$, $ {{\boldsymbol{R}}_v} \triangleq \left( {{\boldsymbol{v}}_R}{\boldsymbol{v}}_R^{\mathrm{T}} +{{\boldsymbol{v}}_I}{\boldsymbol{v}}_I^{\mathrm{T}} \right) $, $ {{\boldsymbol{P}}_v} \triangleq \left( {{\boldsymbol{v}}_R}{\boldsymbol{v}}_I^{\mathrm{T}} - {{\boldsymbol{v}}_I}{\boldsymbol{v}}_R^{\mathrm{T}} \right) $, $ {{\boldsymbol{A}}_{b,R}^{(x)}} \triangleq {\rm{diag}}\left( { {{\rho_b}{{\boldsymbol{\alpha }}_{b,I}}	} } \right) $, $ {{\boldsymbol{A}}_{b,I}^{(x)}} \triangleq {\rm{diag}}\left(\rho_{b}{{\boldsymbol{\alpha }}_{b,R}}\right)$, $ {{\boldsymbol{A}}_{b,R}^{(y)}} \triangleq {\rm{diag}}\left(\eta_{b}{{\boldsymbol{\alpha }}_{b,I}}\right) $, and $ {{\boldsymbol{A}}_{b,I}^{(y)}} \triangleq {\rm{diag}}\left(\eta_{b}{{\boldsymbol{\alpha }}_{b,R}}\right) $, where $\boldsymbol{v}_R$ and $\boldsymbol{v}_I$ are the real part and the imaginary part of $\boldsymbol{v}$, whereas ${{\boldsymbol{\alpha }}_{b,R}}$ and ${{\boldsymbol{\alpha }}_{b,I}}$ are the  real part and the imaginary part of $\boldsymbol{\alpha}(\boldsymbol{t},\boldsymbol{q}_{t},\boldsymbol{q}_{0})$, respectively. Due to space limit, the detailed derivation of gradient vector ${\nabla_{\boldsymbol{t}}}{I_{b,2}}$ is omitted here.

Until now, we have obtained a concave approximation function of $ R_b $ for $\mathcal{P}$3-2-C, which is denoted as ${\hat R_b}\triangleq {\hat I_{b,1}} - {\hat I_{b,2}}$. Similar to $I_{b,2}$, we derive an affine surrogate function of $R_{k}^{\prime}$ to obtain the convex surrogate function of $f_K(\boldsymbol{t})\triangleq\sum_{k=1}^K\left[\mu_k\left(s_k \!- \! R_{k}^{\prime}\right) \!+\! \frac{\rho}{2}\left(s_k \!-\! R_{k}^{\prime}\right)^2\right]$ in \eqref{P3-2-C}. The local affine approximation of $ R_k^{\prime} $ at $\boldsymbol{t}=\boldsymbol{t}^{i_A}$ can be written as 
\begin{align}
	\label{R_k}
	\!{\hat R_k}\triangleq \left.R_{k}^{\prime} \right|_{\boldsymbol{t}={\boldsymbol{t}^{i_A}}} + {\left( {{\nabla _{{{\boldsymbol{t}}^{i_A}}}}{R_{k}^{\prime}}} \right) ^{\mathrm{T}}}\left( {{\boldsymbol{t}} - {{\boldsymbol{t}}^{i_A}}}\! \right).
\end{align}
Denote the real part and the imaginary part of $\boldsymbol{w}$ as $\boldsymbol{w}_R$ and $\boldsymbol{w}_I$, and define $\boldsymbol{\alpha}_{k,R}$ and $\boldsymbol{\alpha}_{k,I}$ as the real part and the imaginary part of $\boldsymbol{\alpha}(\boldsymbol{t},\boldsymbol{q}_{t},\boldsymbol{q}_{k})$, respectively. Then, we define $ {{\boldsymbol{R}}_w} \triangleq \left( {{\boldsymbol{w}}_R}{\boldsymbol{w}}_R^{\mathrm{T}} +{{\boldsymbol{w}}_I}{\boldsymbol{w}}_I^{\mathrm{T}} \right) $, $ {{\boldsymbol{P}}_w} \triangleq \left( {{\boldsymbol{w}}_R}{\boldsymbol{w}}_I^{\mathrm{T}} - {{\boldsymbol{w}}_I}{\boldsymbol{w}}_R^{\mathrm{T}} \right) $, $ {{\boldsymbol{A}}_{k,R}^{(x)}} \triangleq {\rm{diag}}\left( { {{\rho_k}{{\boldsymbol{\alpha }}_{k,I}}	} } \right) $, $ {{\boldsymbol{A}}_{k,I}^{(x)}} \triangleq {\rm{diag}}\left(\rho_{k}{{\boldsymbol{\alpha }}_{k,R}}\right)$, $ {{\boldsymbol{A}}_{k,R}^{(y)}} \triangleq {\rm{diag}}\left(\eta_{k}{{\boldsymbol{\alpha }}_{k,I}}\right) $, and $ {{\boldsymbol{A}}_{k,I}^{(y)}} \triangleq {\rm{diag}}\left(\eta_{k}{{\boldsymbol{\alpha }}_{k,R}}\right) $ with $\rho_k \triangleq \frac{2\pi}{\lambda} \psi_k^x$, $\eta_{k}\triangleq \frac{2\pi}{\lambda} \psi_k^y$. Base on above definitions, ${{\nabla _{{{\boldsymbol{t}}}}}{R_{k}^{\prime}}}$ is given by 
\begin{align}
	\label{eq38}
	{\nabla _{\boldsymbol{t}}}{R_{k}^{\prime}} \!=\!  \frac{1}{{\ln 2}}&\!\left[ \frac{{{\boldsymbol{\mathit{\Gamma}}}_{k,w}^{\left( x \right)}\! \otimes\! {{\boldsymbol{e}}_1} \!+\! {\boldsymbol{\mathit{\Gamma}}}_{k,w}^{\left( y \right)} \!\otimes\! {{\boldsymbol{e}}_2} \!+\! {\boldsymbol{\mathit{\Gamma}}}_{k,v}^{\left( x \right)} \!\otimes\! {{\boldsymbol{e}}_1} \!+\! {\boldsymbol{\mathit{\Gamma}}}_{k,v}^{\left( y \right)} \!\otimes\! {{\boldsymbol{e}}_2}}}{{{\left| {{\boldsymbol{\alpha }}^{\mathrm{H}}}({\boldsymbol{t}},\boldsymbol{q}_{t},\boldsymbol{q}_{k}){\boldsymbol{w}}\right|}^2  \!+\! {\left| {{\boldsymbol{\alpha }}^{\mathrm{H}}}({\boldsymbol{t}},\boldsymbol{q}_{t},\boldsymbol{q}_{k}){\boldsymbol{v}}\right| }^2 \!+\! \sigma _k^{\prime2}}} \right. \notag\\& \left.\qquad\qquad - \frac{{{\boldsymbol{\mathit{\Gamma}}}_{k,v}^{\left( x \right)} \otimes {{\boldsymbol{e}}_1} + {\boldsymbol{\mathit{\Gamma}}}_{k,v}^{\left( y \right)} \otimes {{\boldsymbol{e}}_2}}}{{{\left| {{\boldsymbol{\alpha }}^{\mathrm{H}}}({\boldsymbol{t}},\boldsymbol{q}_{t},\boldsymbol{q}_{k}){\boldsymbol{v}}\right| }^2 + \sigma _k^{\prime2}}} \right]
\end{align}
where $\boldsymbol{\mathit{\Gamma}}_{k,w}^{(x)} \triangleq-\boldsymbol{A}_{k,R}^{(x)}\big(2\boldsymbol{R}_w\boldsymbol{\alpha}_{k,R}+2\boldsymbol{P}_w\boldsymbol{\alpha}_{k,I}\big)+\boldsymbol{A}_{k,I}^{(x)}\big(2\boldsymbol{R}_w\boldsymbol{\alpha}_{k,I}+2\boldsymbol{P}_w^{\mathrm{T}}\boldsymbol{\alpha}_{k,R}\big)$, $\boldsymbol{\mathit{\Gamma}}_{k,w}^{(y)} \triangleq-\boldsymbol{A}_{k,R}^{(y)}\big(2\boldsymbol{R}_w\boldsymbol{\alpha}_{k,R}+2\boldsymbol{P}_w\boldsymbol{\alpha}_{k,I}\big)+\boldsymbol{A}_{k,I}^{(y)}\big(2\boldsymbol{R}_w\boldsymbol{\alpha}_{k,I}+2\boldsymbol{P}_w^{\mathrm{T}}\boldsymbol{\alpha}_{k,R}\big)$, $\boldsymbol{\mathit{\Gamma}}_{k,v}^{(x)} \triangleq-\boldsymbol{A}_{k,R}^{(x)}\big(2\boldsymbol{R}_v\boldsymbol{\alpha}_{k,R}+2\boldsymbol{P}_v\boldsymbol{\alpha}_{k,I}\big)+\boldsymbol{A}_{k,I}^{(x)}\big(2\boldsymbol{R}_v\boldsymbol{\alpha}_{k,I}+2\boldsymbol{P}_v^{\mathrm{T}}\boldsymbol{\alpha}_{k,R}\big)$, and $\boldsymbol{\mathit{\Gamma}}_{k,v}^{(y)} \triangleq-\boldsymbol{A}_{k,R}^{(y)}\big(2\boldsymbol{R}_v\boldsymbol{\alpha}_{k,R}+2\boldsymbol{P}_v\boldsymbol{\alpha}_{k,I}\big)+\boldsymbol{A}_{k,I}^{(y)}\big(2\boldsymbol{R}_v\boldsymbol{\alpha}_{k,I}+2\boldsymbol{P}_v^{\mathrm{T}}\boldsymbol{\alpha}_{k,R}\big)$.

Let ${\hat R_k}$ serve as the surrogate function of $R_k$, and then we have $\hat{f}_K(\boldsymbol{t})\triangleq\sum_{k=1}^K\left[\mu_k\left(s_k \!- \! \hat{R}_k\right) \!+\! \frac{\rho}{2}\left(s_k \!-\! \hat{R}_k\right)^2\right]$, which is a quadratic surrogate function w.r.t $\boldsymbol{t}$. Based on \eqref{I_b,1}, \eqref{I_b,2}, and \eqref{R_k}, for the $ (i_A+1) $-th iteration of the ADMM algorithm, $\mathcal{P}$3-2-C can be reformulated as
\begin{subequations}
	\begin{align}
		\label{P3-2-D}
		\!\!\!\mathcal{P}\text{3-2-D:}
		\min_{\boldsymbol{t}} &\ 	- \!\! \hat{R}_b \!+\!\! \sum_{k=1}^K \!\left[\mu_k \! \left(\! s_k \!- \! \hat{R}_k\right) \!+\! \frac{\rho}{2}\!\left(s_k \!-\! \hat{R}_k\right)^2\right] \\
		\mathrm{s.t.} & \ \boldsymbol{t}_{n}\in\mathcal{C}_n, n =1,2,...,N.
	\end{align}
\end{subequations}
$\mathcal{P}$3-2-D is convex and can be solved by CVX. The details of the ADMM algorithm to solve $\mathcal{P}$3-2 are summarized in Algorithm \ref{algorithm1}.

\begin{algorithm}[t!]
	\caption{ADMM for Solving $\mathcal{P}$3-2 in $ (i+1) $-th AO}
	\label{algorithm1}
	\begin{algorithmic}[1]
		\STATE \textbf{Input}: Set penalty parameter $ \rho>0 $ and convergence threshold $\varepsilon_A$. The solution $\boldsymbol{t}^i$ obtained in $ i $-th iteration of AO serves as the initial solution $\boldsymbol{t}^{i_A}(i_A=0)$ of ADMM. Generate feasible initial solutions $\boldsymbol{s}^0$ and $\boldsymbol{\mu}^0$. Calculate the initial primal residual $\varrho^0=\sqrt{\sum_{k}(s_k^0-R_{k}^{\Delta}(\boldsymbol{t}^0))^2}$ 
		\REPEAT
		\STATE Obtain $\boldsymbol{t}^{i_A+1}$ by solving $\mathcal{P}$3-2-D with given $(\boldsymbol{s}^{i_A},\boldsymbol{\mu}^{i_A})$
		\STATE Obtain $\boldsymbol{s}^{i_A+1}$ by solving \eqref{eq:step2} with given $(\boldsymbol{t}^{i_A+1},\boldsymbol{\mu}^{i_A})$
		\STATE Update $ \boldsymbol{\mu}^{i_A+1} $ according to \eqref{eq:step3}
		\STATE Calculate $\varrho^{i_A+1}= \sqrt{\sum_{k}(s_k^{i_A+1}- R_{k}^{\Delta}(\boldsymbol{t}^{i_A+1}))^2}$
		\IF{$\varrho^{i_A+1}>\varrho^{i_A}/2$}
		\STATE $\rho=\rho\delta_\rho$ with $\delta_\rho$ denoting the adaptive factor of $\rho$
		\ENDIF
		\STATE $i_A=i_A+1$
		\UNTIL{$\varrho^{i_A}<\varepsilon_A $}
	\end{algorithmic}	
\end{algorithm}

\subsection{Overall Algorithm and Computational Complexity}
\begin{algorithm}[tpb]
	\caption{AO Algorithm for Solving $\mathcal{P}0$}
	\label{algorithm2}
	\begin{algorithmic}[1]
		\STATE \textbf{Input}: Set convergence threshold $ \varepsilon $, backtracking factor $\beta$, and $i=0$. Randomly generate feasible initial solutions $\boldsymbol{q}_t^0$,  $\boldsymbol{w}^0$, $\boldsymbol{v}^0$, and $\boldsymbol{t}^0$. 
		\REPEAT
		\STATE Find the most advantageous position $\boldsymbol{q}_k^{\mathrm{opt}}$. Calculate $C_s^i$ based on $(\boldsymbol{q}_t^i,\boldsymbol{w}^i,\boldsymbol{v}^i,\boldsymbol{t}^i,\boldsymbol{q}_k^{\mathrm{opt}})$ according to \eqref{R_s}
		\STATE Given $(\boldsymbol{w}^i,\boldsymbol{v}^i,\boldsymbol{t}^i,\boldsymbol{q}_k^{\mathrm{opt}})$, solve $\mathcal{P}$1-1 to obtain $\boldsymbol{q}_t^{i+1}$ 
		\STATE Given $(\boldsymbol{q}_t^{i+1},\boldsymbol{t}^i,\boldsymbol{q}_k^{\mathrm{opt}})$, solve $\mathcal{P}$2-2 to obtain $ \boldsymbol{W}^{i+1} $ and $ \boldsymbol{V}^{i+1} $. Apply the Gaussian randomization method to recover the rank-one solution $ \boldsymbol{w}^{i+1} $ and $ \boldsymbol{v}^{i+1} $
		\STATE \textbf{FMM}: Given $(\boldsymbol{q}_t^i,\boldsymbol{w}^i,\boldsymbol{v}^i,\boldsymbol{q}_k^{\mathrm{opt}})$, for $n \! =\!1\!\rightarrow\! N$, update $\boldsymbol{t}_n^{i+1}$ by solving P3-1-A
		\\ \textbf{ZMM}: Given $(\boldsymbol{q}_t^i,\boldsymbol{w}^i,\boldsymbol{v}^i,\boldsymbol{q}_k^{\mathrm{opt}})$, obtain $\boldsymbol{t}^{i+1}$ via Algorithm \ref{algorithm1}
		\STATE Calculate $C_s^{i+1} $ based on $(\boldsymbol{q}_t^{i+1}\!,\boldsymbol{w}^{i+1}\!,\boldsymbol{v}^{i+1}\!,\boldsymbol{t}^{i+1}\!,\boldsymbol{q}_k^{\mathrm{opt}})$
		\STATE $i=i+1$
		\UNTIL{$\left|\frac{{C}_s^i-{C}_s^{i-1}}{{C}_s^{i-1}}\right|<\varepsilon $}
	\end{algorithmic}	
\end{algorithm}

The overall algorithm for $\mathcal{P}0$ is summarized in Algorithm \ref{algorithm2}, where the convergence is guaranteed by backtracking method. The computational complexity of Algorithm \ref{algorithm2} is mainly caused by using CVX to solve three subproblems alternatively. First of all, we analyze the computational complexity for solving $\mathcal{P}$2-2. Since $(r_0^\mathrm{opt},r_k^\mathrm{opt})$  can be obtained by analytical expressions \eqref{r_b} and \eqref{r_k}, the complexity of solving $(r_0^\mathrm{opt},r_k^\mathrm{opt})$ is negligible as compared with solving $\boldsymbol{W},\boldsymbol{V}$. With fixed $(r_0^\mathrm{opt},r_k^\mathrm{opt})$, $\mathcal{P}$2-2 has $2N^2$ optimization variables and 3 constraints. Thus, the complexity for solving $\mathcal{P}$2-2 is given by ${\cal O}({L_2}{N^4})$\cite{complexity1}, where $L_2$ denotes the number of iterations of $\mathcal{P}$2-2. Now we turn to $\mathcal{P}$3-1-A, whose number of variables and constraints are 2 and $ (N+3) $, respectively, and thus its computational complexity is ${\cal O}({N^{2.5}})$. On the other hand, the computational complexity of Algorithm \ref{algorithm1} is mainly caused by solving $\mathcal{P}$3-2-D and \eqref{eq:step2}. Note that the number of variables and constraints of $\mathcal{P}$3-2-D are $ N $ and $ 2N $, whereas the number of variables and constraints of \eqref{eq:step2} are $K$ and $K$, respectively. Thus, the computational complexities of $\mathcal{P}$3-2-D and \eqref{eq:step2} are ${\cal O}({N^{3.5}})$ and ${\cal O}({K^{3.5}})$, respectively. Therefore, the complexity of Algorithm \ref{algorithm1} is ${\cal O}\left( {L_3^{{\rm{ZMM}}}\max \left\{ {{N^{3.5}},{K^{3.5}}} \right\}} \right)$, where $L_3^{{\rm{ZMM}}}$ denotes the number of iterations of ADMM. In contrast, the numbers of variables and constraints of $\mathcal{P}$1-1 do not change with $ N $ or $ K $. Thus, the computational complexity of $\mathcal{P}$1-1 is negligible as compared with the other two subproblems as $ N $ and $ K $ increase. According to the foregoing results, the computational complexity of Algorithm \ref{algorithm2} is $ {\cal O}({L_0}{L_2}{N^4}) $ and ${\cal O}\left( {{L_0}\max \left\{ {\left( {{L_2}{N^4}} \right),\left( {L_3^{{\rm{ZMM}}}\max \left\{ {{N^{3.5}},{K^{3.5}}} \right\}} \right)} \right\}} \right)$ under the FMM and the ZMM, respectively, where $L_0$ is the number of iterations required for Algorithm \ref{algorithm2} to converge.

\section{Numerical Results and Discussion}
\label{Section4}
\begin{figure}[t!]  
	\centering
	\subfigure[]{
		\begin{minipage}[b]{0.2\textwidth}
			\includegraphics[width=1.25\textwidth]{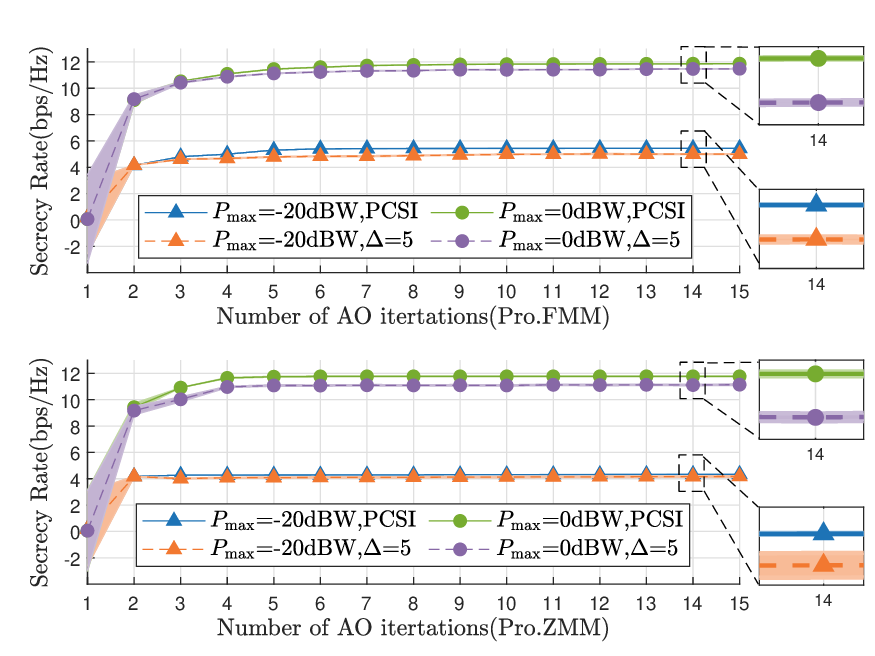}  
		\end{minipage}
	}
	\hfill
	\subfigure[]{
		\begin{minipage}[b]{0.2\textwidth}
			\includegraphics[width=1.1\textwidth]{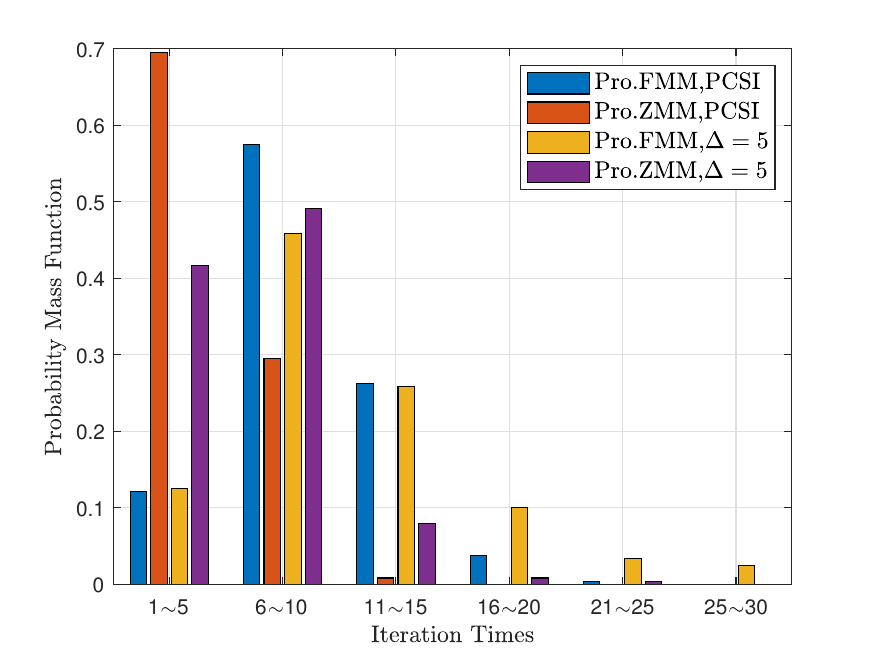}
		\end{minipage}
	}
	\centering
	\caption{Convergence behavior of Algorithm \ref{algorithm2}: (a) MSR versus iterations and (b) relative frequency distribution of iteration times.}
	\label{convergence}
\end{figure}

\begin{figure*}[!t]
	\centering
	\subfigure[]{
		\includegraphics[scale=0.3]{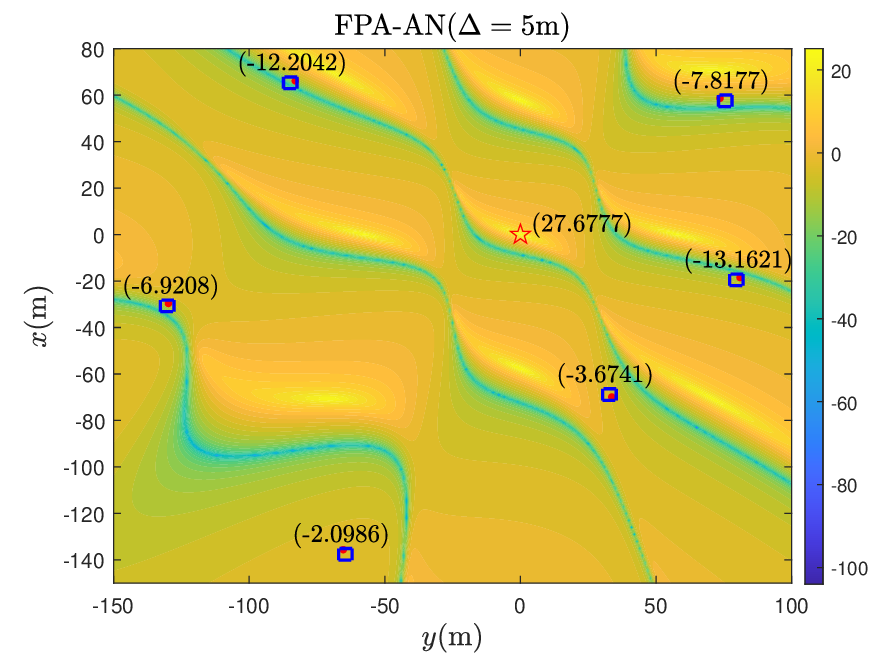}}
	\hfil
	\subfigure[]{
		\includegraphics[scale=0.3]{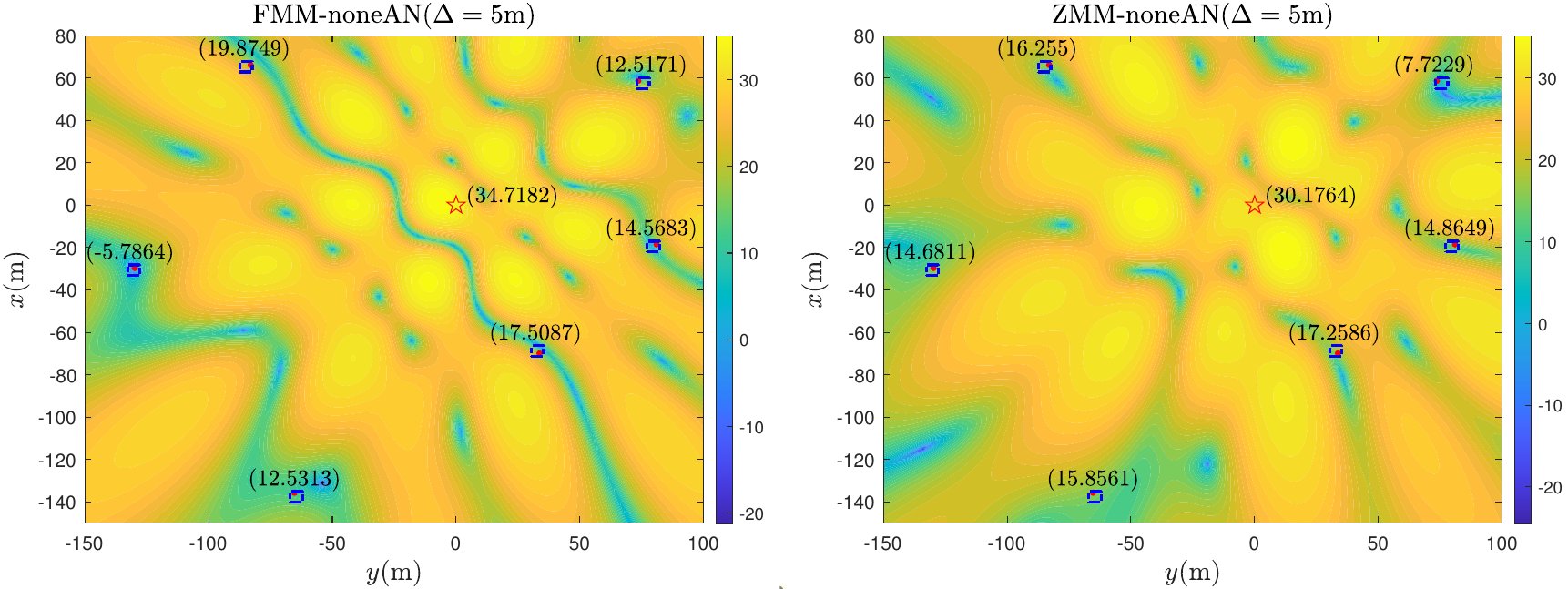}}
	\\
	\subfigure[]{
			\includegraphics[scale=0.3]{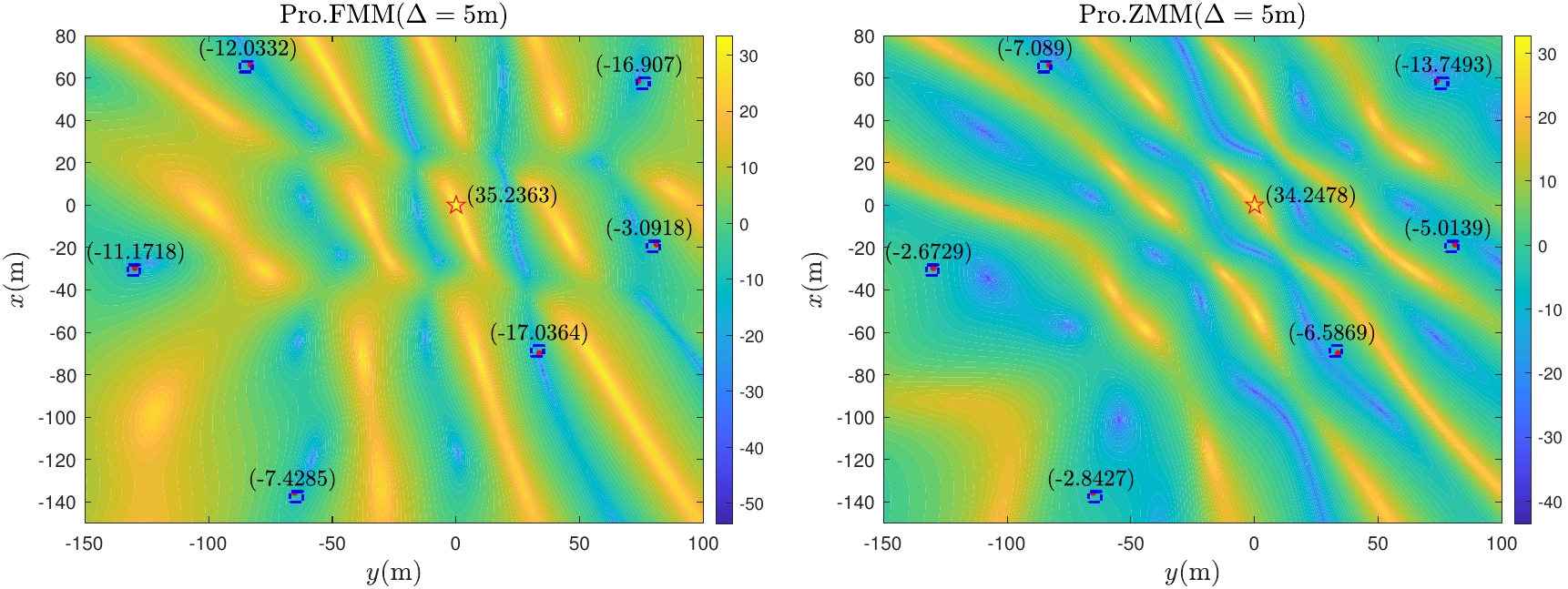}}
		\hfill\hfil
	\subfigure[]{
		\includegraphics[scale=0.3]{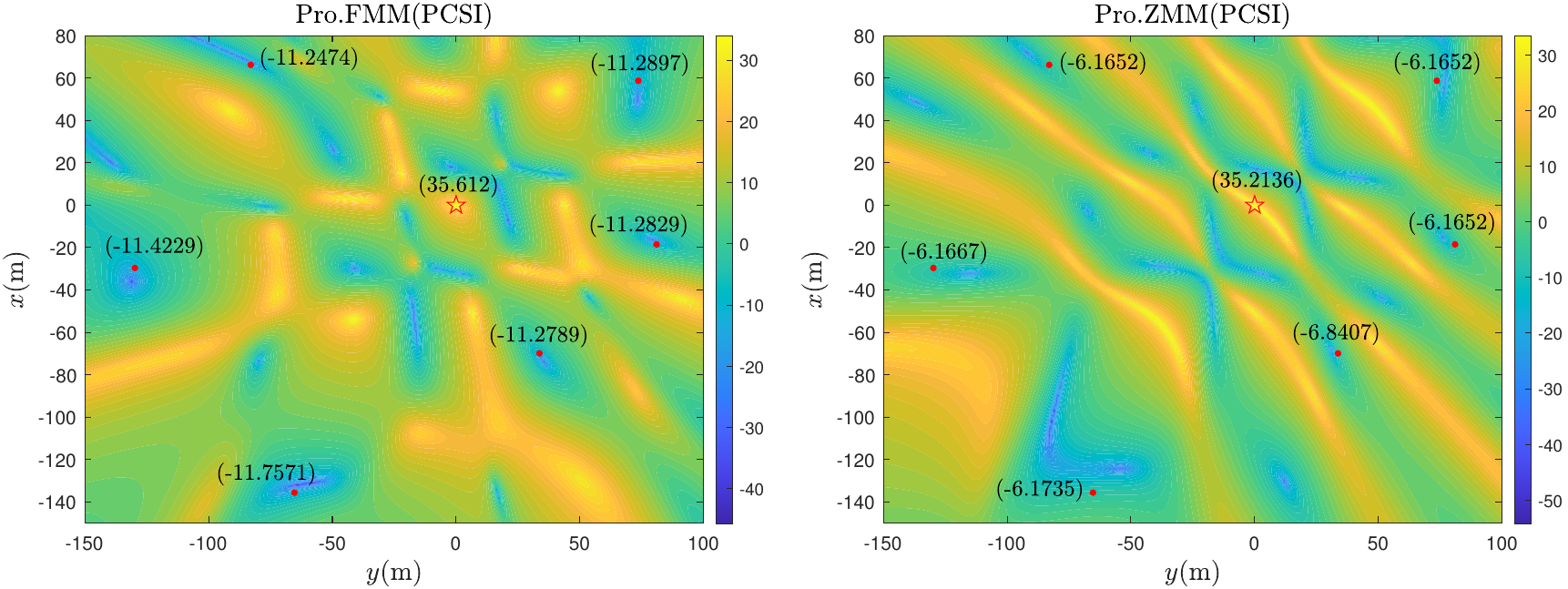}}
	\caption{Comparisons of the SINR pattern of (a) FPA-AN scheme, (b) FMM-noneAN and ZMM-noneAN schemes, and (c) Pro. FMM and Pro. ZMM schemes under $\Delta=5$ m; (d) Pro. FMM and Pro. ZMM schemes under PCSI.}
	\label{SINR}
\end{figure*}

This section presents representative numerical results to evaluate the performance of proposed schemes. The position of Bob is set at $ \boldsymbol{q}_0=[0,0]^\mathrm{T} $ in meter. The uncertain region of the $k$-th Eve is set as $\mathcal{A}_k=\left\{\left.[x_k, y_k]^{\mathrm{T}}\right| x_{k} \in [\hat{x}_k -\frac{\Delta}{2},\hat{x}_k + \frac{\Delta}{2}],{y_k} \in [\hat{y}_k - \frac{\Delta}{2},\hat{y}_k + \frac{\Delta}{2}]\right\}$, where $ [\hat{x}_k,\hat{y}_k]^{\mathrm{T}}\triangleq \hat{\boldsymbol{q}}_k $ is the center of the suspicious region of the $k$-th Eve. $\{\hat{\boldsymbol{q}}_k\}_{k=1}^K$ are chosen in sequence from a randomly generated set $\hat{\mathcal{S}}=$\{[57.5734, 75.5166]$ ^{\mathrm{T}} $, [-19.4950, 79.6046]$ ^{\mathrm{T}} $, [-68.8583, 32.8196]$ ^{\mathrm{T}} $, [-137.6302, -64.5212]$ ^{\mathrm{T}} $, [-30.6881, -130.2107]$ ^{\mathrm{T}} $, [65.3468, -84.8546]$ ^{\mathrm{T}} $\} in meter. 
Unless otherwise specified, the system parameters are set as $N=4$, $K=6$, $\sigma_b^2=\sigma_{k}^2=-105$ dBW\cite{noise}, $\lambda=0.01$ m, $ d_{\min}=\lambda/2 $, $C=4\lambda$, $\xi_0=10^{-3}$, $\alpha_{b}=\alpha_{k}=2.2$, $H_0=100$ m, $x_t^{\min } = y_t^{\min } =  - 150$ m, $x_t^{\max } = y_t^{\max } = 150$ m, $\varepsilon=\varepsilon_A=\varepsilon_D=10^{-3}$, $\beta=0.8$, and $\Delta=5$ m. Without loss of generality, the accurate position of the $ k $-th Eve is randomly generated within uncertain region $\mathcal{A}_k$.
Each result is obtained by averaging over $ 10^3 $ independent channel realizations. The two proposed schemes (\textbf{Pro. FMM} and \textbf{Pro. ZMM}) are compared with the following benchmark schemes.
\begin{itemize}
	\item \textbf{FPA-noneAN}\cite{FPA-noneAN}: The AAV adopts FPA-based uniform planar array (UPA) with antenna spacings $\lambda/2$, and the AN is not utilized.
	\item \textbf{FPA-AN}\cite{9TIFS24}: The AAV adopts FPA-based UPA, and the AN is incorporated in transmit signal.
	\item \textbf{FMM-noneAN}\cite{3_Secrecy_LSP24}: The AAV adopts 2D FA array under the FMM, and the AN is not utilized.
	\item \textbf{ZMM-noneAN}: The AAV adopts 2D FA array under the ZMM, and the AN is not utilized. The proposed Algorithm \ref{algorithm2} is applied.
\end{itemize}

Figure \ref{convergence} shows the convergence behavior of the AO algorithm under both perfect CSI (PCSI) ($\Delta\!=\!0$ m) and imperfect CSI ($\Delta\!=\!5$ m) conditions. Specifically, in Figure \ref{convergence}(a), each line represents the average of the MSRs $ ({R_s}\! \triangleq \!\max\nolimits_k {\left(R_{0}-R_{{k}} \right)})$ obtained from different random initializations, denoted as $\mu_\mathrm{MSR}$. The height of each shaded region is twice the standard deviation of $R_s/{\mu_\mathrm{MSR}}$.
As can be observed, the MSRs exhibit more concentrated distribution under PCSI compared to $ \Delta=5 $ m, but Algorithm \ref{algorithm2} still exhibits superior stability under $ \Delta=5 $ m. Moreover, Figure \ref{convergence}(b) shows the relative frequency distribution of the required iterations for the AO algorithm to converge, where we set $ P_{\max}=0$ dBW. For the ZMM, the relative frequency of reaching convergence within 20 iterations is $ 1 $ under PCSI and $ 0.99 $ under $ \Delta=5 $ m. For the FMM, the relative frequency of reaching convergence within 20 iterations is $ 0.995 $ under PCSI and $ 0.94 $ under $ \Delta=5 $ m. These observations mean that Algorithm \ref{algorithm2} exhibits a rapid convergence rate.

Figure \ref{SINR} illustrates the SINR (in dB) of the proposed schemes and the benchmarks at different reception positions based on once realization, where we set $ P_{\max}=0 $ dBW. The star point represents position of Bob, whereas the solid points represent accurate positions of Eves, and the dotted boxes denote the uncertain regions of Eves. Several observations are drawn from the figure: 1) Compared with the proposed schemes (shown in Figure \ref{SINR}(c)), the FPA-AN scheme (shown in Figure \ref{SINR}(a)) sacrifices 7$\sim$8 dB SINR at Bob in order to suppress Eves, which validates that FA can achieve a better trade-off between reliability and security than FPA. 2) Pro. FMM and Pro. ZMM (shown in Figure \ref{SINR}(c)) significantly outperform FMM-noneAN and ZMM-noneAN (shown in Figure \ref{SINR}(b)), which indicates the necessity of AN necessity for security performance enhancement. 3) In the case of PCSI (shown in Figure \ref{SINR}(d)), the SINR of the proposed schemes are approximately the same at each wiretap position, which means the MSR is effectively optimized and verifies the effectiveness of the proposed algorithm. In contrast, under imperfect CSI (shown in Figure \ref{SINR}(c)), the limited ECSI prevents the transmitter from uniformly suppressing eavesdroppers, resulting in unavoidable SINR variance among eavesdroppers. Even though, as will be shown later, the proposed schemes can still effectively enhance the MSR.

\begin{figure}[t!]
	\centering
	\includegraphics[width=3in]{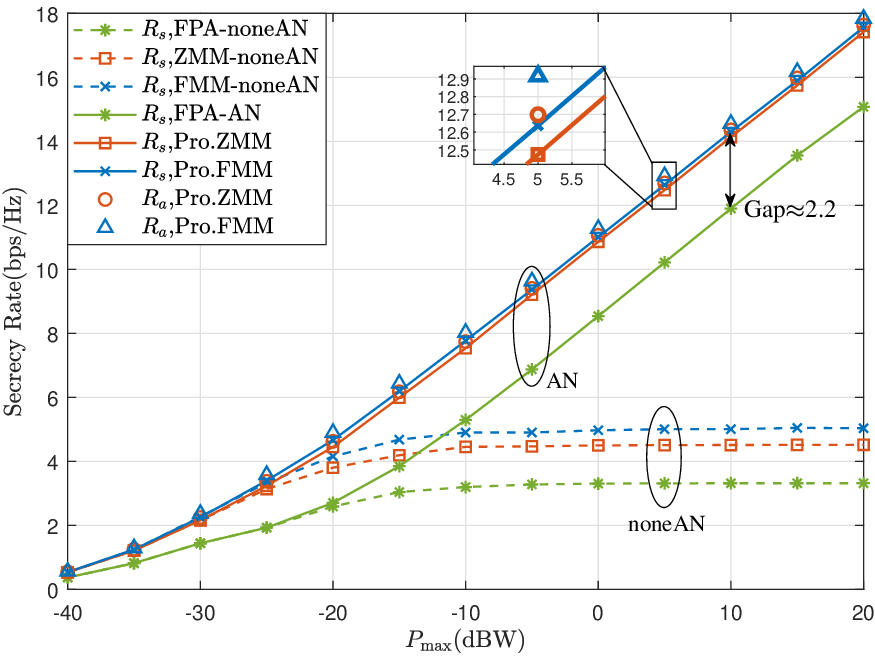}
	\caption{Comparisons of the SRs of different schemes.}
	\label{SR_P}
\end{figure}

Figure \ref{SR_P} shows the MSR of different schemes and the average SR $ ({R_a} \!\triangleq\! \sum\nolimits_k {\left(R_{0}\!-\! R_{{k}} \right)} /K)$ of proposed schemes under $\Delta \! =\!5$ m. Three observations are drawn from the figure: 1) Regardless of whether AN is adopted or not, all FA schemes outperform their FPA counterparts across whole transmit power range, which benefits from the additional DoF provided by APV optimization. 2) 
When $ P_{\max} $ exceeds approximately $ -12 $ dBW, and the SR of FPA-AN scheme is larger than both FMM-noneAN and ZMM-nonAN schemes. This means that the AN provides higher confidentiality gains as compared with the FA technique when the transmit power is sufficient. 3) The $ R_s $ and $ R_a $ of the proposed schemes match well, which verifies that the proposed algorithm can effectively maximize the MSR.

\begin{figure}[t!]
	\centering
	\includegraphics[width=3in]{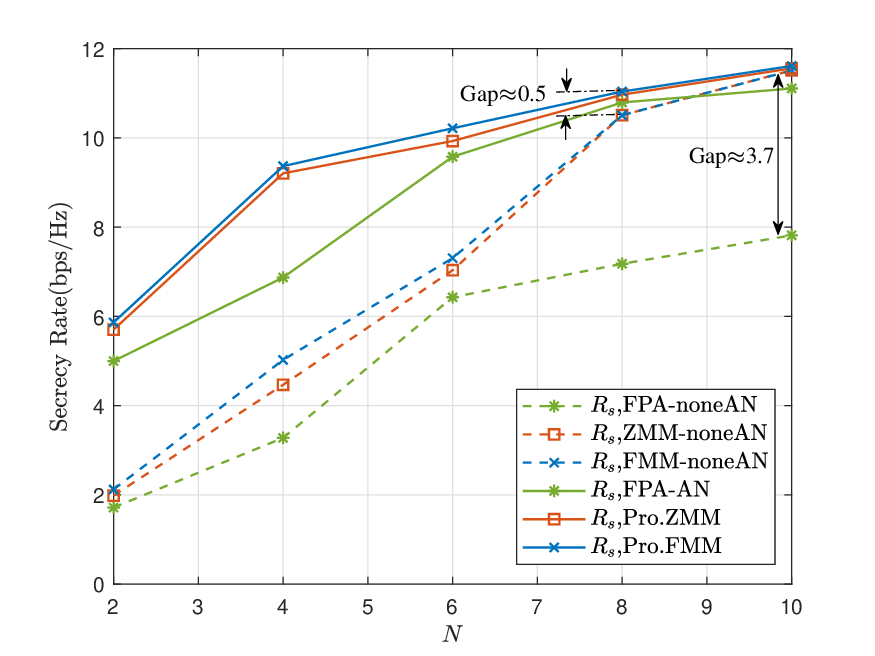}
	\caption{SRs versus the number of antennas for FA-aided schemes.}
	\label{SR_N}
\end{figure}

\begin{figure}[t!]
	\centering
	\includegraphics[width=3in]{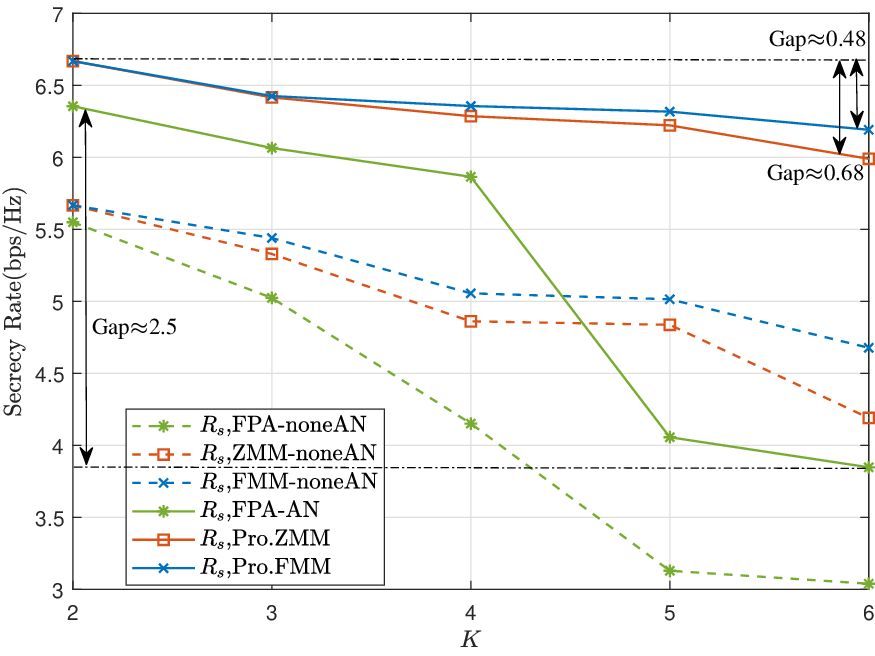}
	\caption{Impacts of the number of eavesdroppers on the SRs.}
	\label{SR_K}
\end{figure}

Figure \ref{SR_N} shows the impacts of the number of FAs on the SR, where $ P_{\max} = -5 $ dBW and $\Delta \! =\!5$ m. As can be seen, the SR increases monotonically with $ N $ for all schemes. This is because more antennas can bring more spatial DoF and higher array gain. As for FA schemes, the performance gain of the AN becomes small when the number of transmit antenna exceeds $ N=8 $, which implies that the sufficient spatial DoF from FA alleviates the dependence on AN. Moreover, note that when $N=10$, the SRs of FA-noneAN schemes slightly outperform those of the FPA-AN scheme. This indicates that the DoF brought by FA is more advantageous than the auxiliary gain provided by AN when the number of transmit antennas is sufficient.

Figure \ref{SR_K} illustrates the impacts of the number of eavesdroppers on the SR, where $ P_{\max} = -15 $ dBW and $\Delta=5$ m. It can be observed that the MSRs of FPA schemes degrade more rapidly than their FA counterparts, regardless of whether AN is adopted or not, which means that FA can effectively resist the multiple eavesdropping scenario. This is because, instead of maintaining fixed spacing as FPA, FAs can be adjusted to be more dispersed, which will result in more null lobes and thus more eavesdroppers can be suppressed simultaneously \cite{grating_lobe}. In contrast, the FPA scheme does not have sufficient DoF to effectively suppress signals at all Eves when the number of transmit antennas is less than that of Eves.

\begin{figure}[t!]
	\centering
	\includegraphics[width=2.78in]{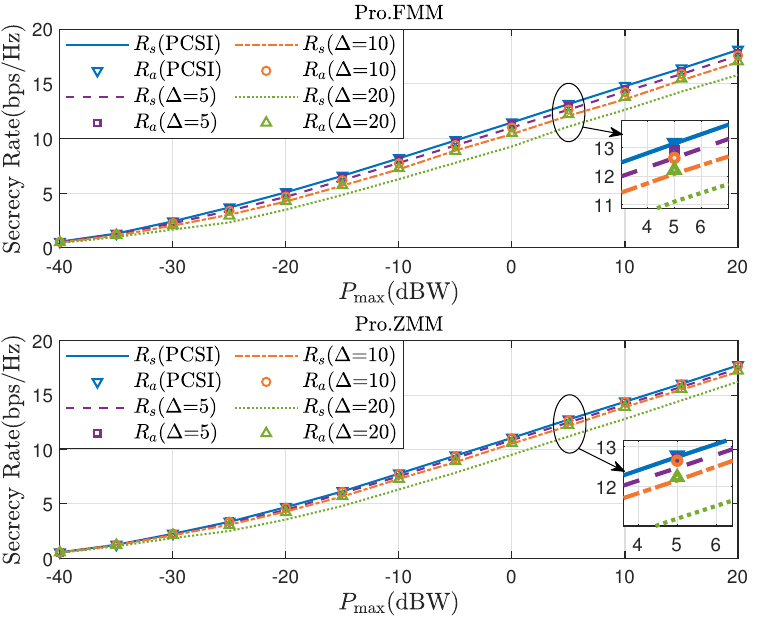}
	\caption{Impacts of the ECSI uncertainty on the SRs.}
	\label{robust}
\end{figure}

To further illustrate the robustness of the proposed schemes, the impact of ECSI uncertainty on security performance is evaluated in Figure \ref{robust}. As expected, the SR decreases as the ECSI uncertainty range $\Delta$ increases. Compared with PCSI, the loss of MSR is approximately $ 0.4 $ bps/Hz, $ 1 $ bps/Hz, and $ 2 $ bps/Hz for $\Delta=5$ m, $\Delta=10$ m, and $\Delta=20$ m, respectively. In addition, the gap between $ R_s $ and $ R_a $ is increasing as ECSI uncertainty becomes lager, which is because that the larger uncertainty makes it difficult to achieve a balanced deep fading at different eavesdroppers' positions. Despite this, the MSR remains a high level even when $\Delta=20$ m, which verifies the superior robustness of the proposed schemes.

\section{Concluding Remarks}
\label{Section5}
This paper investigated a FA-AN-assisted AAV secure communication system under imperfect ECSI. Specifically, a practical scenario where the transmitter only knows the regions where eavesdroppers possibly exist was considered. Then, a robust optimization algorithm was proposed to maximize the worst-case MSR by jointly designing the AAV deployment, precoders, and antenna positions. In particular, two efficient algorithms to solve the subproblems corresponding to the two antenna movement modes were designed. Numerical results revealed that both the FA and the AN techniques can effectively enhance the security. In particular, AN is more important as compared with FA when transmit power is sufficient. Moreover, the ZMM is superior than the FMM when the number of eavesdroppers is large. Furthermore, the superior robustness of the proposed schemes has been verified within a large ECSI uncertainty range.

\begin{appendices}
	\section{Proof of Proposition 1}
	\label{Appendix_A}
	Since ${\Omega _k}$  is the convex hull of ${\Lambda _k}$, we have ${\Lambda _k} \subseteq {\Omega _k},\forall k \in {\cal K}$, which implies
	\begin{align}
		\label{hull_1}
		\!\!\!\max_{\boldsymbol{A}_k\in\Lambda_k}\boldsymbol{w}^{\mathrm{H}} \!\! \boldsymbol{A}_k\boldsymbol{w}\!-\!\zeta\boldsymbol{v}^{\mathrm{H}} \!\!\boldsymbol{A}_k\boldsymbol{v}\!\leq\!\!\max_{\boldsymbol{A}_k\in\Omega_k}\boldsymbol{w}^{\mathrm{H}} \!\!\boldsymbol{A}_k\boldsymbol{w}\!-\!\zeta\boldsymbol{v}^{\mathrm{H}} \!\! \boldsymbol{A}_k\boldsymbol{v}.
	\end{align}
	In addition, according to \eqref{convex_hull}, for $\forall\boldsymbol{A}_k\in\Omega_k$, the objective function can be decomposed as
	\begin{align}
		\!\!\!\boldsymbol{w}^{\mathrm{H}} \!\boldsymbol{A}_k\boldsymbol{w}\!-\!\zeta\boldsymbol{v}^{\mathrm{H}} \!\boldsymbol{A}_k\boldsymbol{v}=\!\!\sum_{i=1}^{S_k}\tau_{k,i}\!\left(\boldsymbol{w}^{\mathrm{H}} \!\boldsymbol{A}_{k,i}\boldsymbol{w}\!-\!\zeta\boldsymbol{v}^{\mathrm{H}} \!\boldsymbol{A}_{k,i}\boldsymbol{v}\right).
	\end{align}
	Since $\sum\limits_{i = 1}^{{S_k}} {{\tau _{k,i}}}  = 1$ and ${\tau _{k,i}} \ge 0$ hold, there must exist a ${{\boldsymbol{A}}_{k,i^\prime}} \in {\Lambda _k}$ satisfying
	\begin{align}
		\boldsymbol{w}^{\mathrm{H}}\boldsymbol{A}_k\boldsymbol{w}-\zeta \boldsymbol{v}^{\mathrm{H}}\boldsymbol{A}_k\boldsymbol{v}\leq \boldsymbol{w}^{\mathrm{H}}\boldsymbol{A}_{k,i^{\prime}}\boldsymbol{w}-\zeta \boldsymbol{v}^{\mathrm{H}}\boldsymbol{A}_{k,i^{\prime}}\boldsymbol{v}
	\end{align}
	which implies the existence of $\boldsymbol{A}_{k,i^{\prime}}\in\Lambda _k$  such that
	\begin{align}
		\label{hull_2}
		\max_{\boldsymbol{A}_{k,i^{\prime}}\in\Lambda_k}\boldsymbol{w}^{\mathrm{H}} \!\! \boldsymbol{A}_{k,i^{\prime}}\boldsymbol{w}\!-\!\zeta\boldsymbol{v}^{\mathrm{H}} \!\!\boldsymbol{A}_{k,i^{\prime}}\boldsymbol{v}\!\ge\!\!\max_{\boldsymbol{A}_k\in\Omega_k}\boldsymbol{w}^{\mathrm{H}} \!\!\boldsymbol{A}_k\boldsymbol{w}\!-\!\zeta\boldsymbol{v}^{\mathrm{H}} \!\! \boldsymbol{A}_k\boldsymbol{v}.
	\end{align}
	Combining \eqref{hull_1} and \eqref{hull_2} yields the equivalence stated in Proposition 1, which completes the proof.
		
	\section{Derivations of \texorpdfstring{$\nabla_{\boldsymbol{q}_{t}}R_{b}$}{}}
	\label{Appendix_B}
	The gradient vector of composite function $ R_b $ can be expressed as
	\begin{align}
		\label{R_qt}
		\nabla_{\boldsymbol{q}_t}R_b=\frac{\nabla_{\boldsymbol{q}_t}h_{w}+\nabla_{\boldsymbol{q}_t}h_{v}+\sigma_b^2\nabla_{\boldsymbol{q}_t}h_d}{\ln2\left(h_{w}(\boldsymbol{q}_t)+h_{v}(\boldsymbol{q}_t)+\sigma_b^2h_d\left(\boldsymbol{q}_t\right)\right)}\notag\\
		-\frac{\nabla_{\boldsymbol{q}_t}h_{v}+\sigma_b^2\nabla_{\boldsymbol{q}_t}h_d}{\ln2\left(h_v(\boldsymbol{q}_t)+\sigma_b^2h_d\left(\boldsymbol{q}_t\right)\right)}.
	\end{align}
	To proceed,  $h_{w}(\boldsymbol{q}_t)$ is rewritten as
	\begin{align}
		h_{w}&(\boldsymbol{q}_{t}) =\left|\sum_{n=1}^{N}\left|w_{n}\right|e^{-j\frac{2\pi}{\lambda}\left(x_{n}\frac{x_0-x_{t}}{d_{0}}+y_{n}\frac{y_{0}-y_{t}}{d_{0}}\right)+j\angle w_{n}}\right|^{2} \\
		& =\sum_{n=1}^{N}\sum_{m=1}^{N}|w_{n}w_{m}|\cos\left(\frac{2\pi}{\lambda}\left(\frac{\left(x_{n}-x_{m}\right)\left(x_0-x_{t}\right)}{d_0}\right.\right. \notag\\ &\left.\left. +\frac{\left(y_{n}-y_{m}\right)\left(y_0-y_{t}\right)}{d_0}\right)-\left(\angle w_{n}-\angle w_{m}\right)\right).
	\end{align}
	Then, $\nabla_{\boldsymbol{q}_{t}}h_{w}\triangleq\left[\frac{\partial h_{w}(\boldsymbol{q}_{t})}{\partial x_{t}},\frac{\partial h_{w}(\boldsymbol{q}_{t})}{\partial y_{t}}\right]^{\mathrm{T}}$ can be derived as
	\begin{align}
		\label{hw_x}
		\frac{\partial h_{w}\left(\boldsymbol{q}_t\right)}{\partial x_t}\!=&\frac{2\pi}{\lambda d^2_0}\sum_{n=1}^N \!\sum_{m=1}^N \!\!\left|w_nw_m\right|\!\sin\!\left(\!\frac{2\pi}{\lambda}\vartheta_{n,m}\!-\!\left(\angle w_n \!\!-\!\!\angle w_m\right)\!\right)\notag\\ &\quad\times\left(\left(x_n \! -\! x_m\right)d_0\!+\!\vartheta_{n,m}\left(x_t \! -\! x_0\right)\right) \\
		\frac{\partial h_{w}\left(\boldsymbol{q}_t\right)}{\partial y_t}\!=&\frac{2\pi}{\lambda{d_0^2}}\sum_{n=1}^N \!\sum_{m=1}^N \!\!|w_nw_m|\!\sin\!\left(\!\frac{2\pi}{\lambda}\vartheta_{n,m} \!-\!\left(\angle w_n \!\!-\!\!\angle w_m\right)\!\right)\notag\\ &\qquad\times\left( \left(y_n \!- \! y_m\right)d_0\!+\!\vartheta_{n,m}\left(y_t \!- \! y_0\right)\right)
	\end{align}
	where ${\vartheta _{n,m}} = \frac{{\left( {{x_n} - {x_m}} \right)\left( {{x_0} - {x_t}} \right) + \left( {{y_n} - {y_m}} \right)\left( {{y_0} - {y_t}} \right)}}{{{d_0}}}$. The gradient vector $\nabla_{\boldsymbol{q}_{t}}h_{v}$ can be derived by a similar process as $\nabla_{\boldsymbol{q}_{t}}h_{w}$.
	In addition, the gradient vector $\nabla_{\boldsymbol{q}_{t}}h_{d}\triangleq\left[\frac{\partial h_{d}(\boldsymbol{q}_{t})}{\partial x_{t}},\frac{\partial h_{d}(\boldsymbol{q}_{t})}{\partial y_{t}}\right]^{\mathrm{T}}$ can be derived as
	\begin{align}
		\label{hd}
		\!\!\nabla_{\boldsymbol{q}_{t}}h_{d} \!=\! {\left[ {\frac{{{\alpha _0}\left( {{x_t} \!-\! {x_0}} \right){d_0}^{{\alpha _0} - 2}}}{{{\zeta _0}}},\frac{{{\alpha _0}\left( {{y_t} \!-\! {y_0}} \right){d_0}^{{\alpha _0} - 2}}}{{{\zeta _0}}}} \right]^{\mathrm{T}}}.
	\end{align}
	Finally, $\nabla_{\boldsymbol{q}_t}R_b$ is obtained by submitting \eqref{hw_x}-\eqref{hd} into \eqref{R_qt}.

	\section{Proof of Lemma 2}
	\label{Appendix_C}
	The second-order Taylor expansion of $g_b(\boldsymbol{t}_n)$ at  $\boldsymbol{t}_n=\boldsymbol{t}_n^i$ can be written as
	\begin{align}
		g_{b}(\boldsymbol{t}_{n})&=g_{b}\left(\boldsymbol{t}_{n}^{i}\right)+\left(\nabla_{\boldsymbol{t}_{n}^{i}}g_{b}\right)^{\mathrm{T}}\left(\boldsymbol{t}_{n}-\boldsymbol{t}_{n}^{i}\right)\notag\\&+\frac{1}{2}\left(\boldsymbol{t}_{n}-\boldsymbol{t}_{n}^{i}\right)^{\mathrm{T}}\left(\nabla_{\boldsymbol{t}_{n}}^{2}g_{b}\right)\left(\boldsymbol{t}_{n}-\boldsymbol{t}_{n}^{i}\right)+R_{2}\left(\boldsymbol{t}_{n}\right)
	\end{align}
	where $R_{2}(\boldsymbol{t}_{n})$ is the remainder term representing higher-order infinitesimals. ${\nabla _{{{\boldsymbol{t}}_n}}}{g_{b}} = {\left[ {\frac{{\partial {g_{b}}\left( {{{\boldsymbol{t}}_n}} \right)}}{{\partial {x_n}}},\frac{{\partial {g_{b}}\left( {{{\boldsymbol{t}}_n}} \right)}}{{\partial {y_n}}}} \right]^{\mathrm{T}}}$ is the gradient vector of $g_{b}(\boldsymbol{t}_{n})$ at ${\boldsymbol{t}_n} = {\left[ {{x_n},{y_n}} \right]^{\mathrm{T}}}$, which is given by
	\begin{align}
		\frac{{\partial {g_{b}}\left( {{{\boldsymbol{t}}_n}} \right)}}{{\partial {x_n}}} =  - \left| {{\Xi _{wv}}} \right|{\rho _b}\sin \left( {{\rho _b}{x_n} + {\eta _b}{y_n} - \angle {\Xi _{wv}}} \right)\\
		\frac{{\partial {g_{b}}\left( {{{\boldsymbol{t}}_n}} \right)}}{{\partial {y_n}}} =  - \left| {{\Xi _{wv}}} \right|{\eta _b}\sin \left( {{\rho _b}{x_n} + {\eta _b}{y_n} - \angle {\Xi _{wv}}} \right).
	\end{align}
	The Hessian matrix $\nabla _{{{\boldsymbol{t}}_n}}^2{g_{b}} = \left[ \begin{array}{l}
		\frac{{\partial {g_{b}}\left( {{{\boldsymbol{t}}_n}} \right)}}{{\partial {x_n}\partial {x_n}}},\frac{{\partial {g_{b}}\left( {{{\boldsymbol{t}}_n}} \right)}}{{\partial {x_n}\partial {y_n}}}\\
		\frac{{\partial {g_{b}}\left( {{{\boldsymbol{t}}_n}} \right)}}{{\partial {y_n}\partial {x_n}}},\frac{{\partial {g_{b}}\left( {{{\boldsymbol{t}}_n}} \right)}}{{\partial {y_n}\partial {y_n}}}
	\end{array} \right]$ can be specified as
	\begin{align}
		\frac{{\partial {g_{b}}\left( {{{\boldsymbol{t}}_n}} \right)}}{{\partial {x_n}\partial {x_n}}} =  - \left| {{\Xi _{wv}}} \right|{\rho _b}^2\cos \left( {{\rho _b}{x_n} + {\eta _b}{y_n} - \angle {\Xi _{wv}}} \right)\\
		\frac{{\partial {g_{b}}\left( {{{\boldsymbol{t}}_n}} \right)}}{{\partial {x_n}\partial {y_n}}} =  - \left| {{\Xi _{wv}}} \right|{\eta _b}{\rho _b}\cos \left( {{\rho _b}{x_n} + {\eta _b}{y_n} - \angle {\Xi _{wv}}} \right)\\
		\frac{{\partial {g_{b}}\left( {{{\boldsymbol{t}}_n}} \right)}}{{\partial {y_n}\partial {x_n}}} =  - \left| {{\Xi _{wv}}} \right|{\eta _b}{\rho _b}\cos \left( {{\rho _b}{x_n} + {\eta _b}{y_n} - \angle {\Xi _{wv}}} \right)\\
		\frac{{\partial {g_{b}}\left( {{{\boldsymbol{t}}_n}} \right)}}{{\partial {y_n}\partial {y_n}}} =  - \left| {{\Xi _{wv}}} \right|{\eta _b}^2\cos \left( {{\rho _b}{x_n} + {\eta _b}{y_n} - \angle {\Xi _{wv}}} \right)
	\end{align}
	Based on the definitions of Euclidean norm and Frobenius norm of a matrix, we have
	\begin{align}
		& \left\|\nabla_{t_{n}}^{2}g_{b}\right\|_{2}^{2}\leq\left\|\nabla_{t_{n}}^{2}g_{b}\right\|_\mathrm{F}^{2} \notag\\
		& =\left(\frac{\partial g_{b}\left(\boldsymbol{t}_{n}\right)}{\partial x_{n}\partial x_{n}}\right)^{2}\!\!\!\!+\! \left(\frac{\partial g_{b}\left(\boldsymbol{t}_{n}\right)}{\partial x_{n}\partial y_{n}}\right)^{2}\!\!\!\!+\!\left(\frac{\partial g_{b}\left(\boldsymbol{t}_{n}\right)}{\partial y_{n}\partial x_{n}}\right)^{2}\!\!\!\!+\!\left(\frac{\partial g_{b}\left(\boldsymbol{t}_{n}\right)}{\partial y_{n}\partial y_{n}}\right)^{2} \notag\\
		& \leq\left(\left|\Xi_{w\nu}\right|\rho_{b}{}^{2}\right)^{2}+2\left(\left|\Xi_{w\nu}\right|\eta_{b}\rho_{b}\right)^{2}+\left(\left|\Xi_{w\nu}\right|\eta_{b}{}^{2}\right)^{2} \notag\\
		& =\left(\left|\Xi_{w\nu}\right|\left({\rho_{b}}^{2}+{\eta_{b}}^{2}\right)\right)^{2}
	\end{align}
	Since ${\left\| {\nabla _{{{\boldsymbol{t}}_n}}^2{g_{b}}} \right\|_2}{{\boldsymbol{I}}_2}\succcurlyeq\nabla _{{{\boldsymbol{t}}_n}}^2{g_{b}}$, we can set
	\begin{align}
		{\gamma _{b,n}} = \left| {{\Xi _{b,wv}}} \right|\left( {{\rho _b}^2 + {\eta _b}^2} \right)
	\end{align}
	which satisfies $\gamma_{b,n}{{\boldsymbol{I}}_2}\succcurlyeq\nabla _{{{\boldsymbol{t}}_n}}^2{g_{b}}$. Based on Lemma 12 of\cite{MM}, we have
	\begin{align}
		g_{b}\left(\boldsymbol{t}_{n}\right)\geq& g_{b}\left(\boldsymbol{t}_{n}^{i}\right)+\nabla_{t_{n}^{i}}^{\mathrm{T}}g_{b}\left(\boldsymbol{t}_{n}\right)\left(\boldsymbol{t}_{n}-\boldsymbol{t}_{n}^{i}\right)\notag\\&\quad-\frac{\gamma_{n}}{2}\left(\boldsymbol{t}_{n}-\boldsymbol{t}_{n}^{i}\right)^{\mathrm{T}}\left(\boldsymbol{t}_{n}-\boldsymbol{t}_{n}^{i}\right)\triangleq\tilde{g}_{b}\left(\boldsymbol{t}_{n}\right)
	\end{align}
	which completes the proof.
		
\end{appendices}

\bibliographystyle{IEEEtran}
\bibliography{ref}
\end{document}